\documentclass[12pt]{article}
\usepackage{amsfonts}
\begin{document}
\begin{flushright}
arXiv:0812.2889 [hep-th], \\ %December 15, 2008 \\  V3: April 20, 2009 \\
 JHEP 04, 085 (2009)
% Printed \today
\end{flushright}
\vspace{3cm}

\begin{center}
 {\bf \Large On superembedding approach to type IIB 7--branes}

\vskip 2cm

 {\bf Igor A. Bandos}

\vskip 1.5cm

{\it Ikerbasque, the Basque Science Foundation, and Department of
Theoretical Physics, \\ The University of the Basque Country
 (EHU/UPV), P.O. Box 644, 48080 Bilbao, Spain}

and \\ {\it Institute for Theoretical Physics, NSC KIPT,
 61108 Kharkov, Ukraine}

\vskip 1.5cm

{\bf Abstract}
\end{center}
In search for a dynamical description of Q7-branes, which were known
as solutions of supergravity equations and then conjectured to be
dynamical objects of type IIB string theory, we study the
superembedding description of 7-branes in curved type IIB
supergravity superspace. With quite minimal and natural assumptions
we have found that there is no place for Q7-branes as dynamical
branes in superembedding approach. As Q7--brane was also considered
as a bound state of two SD7-branes (this is to say of two 7-branes
related to the D7--brane by different SL(2) transformation), our
study  might give implications for the old-standing problem of the
covariant and supersymmetric description of multiple Dirichlet
$p$-brane systems.

\bigskip

\thispagestyle{empty}

\newpage
{\small \tableofcontents }

\section{Introduction }

Some times ago a new type of 7-brane solutions of type IIB
supergravity was described in \cite{Eric+06sD7}. Latter this
solutions were called Q-branes \cite{Dima+Eric07} and it was
conjectured \cite{Eric+06aD7,Dima+Eric07} that they may  correspond
to a supersymmetric extended object, also called Q7-brane, in the
same manner as the M2-brane solution of D=11 supergravity
\cite{Duff:1990xz} is associated with the D=11 supermembrane action
of \cite{BST1987}. A support for such a conjecture was the existence
of a triplet of eight-form fields in D=10 type IIB supergravity,
which had been found for the first time in \cite{D'A+L+T98}, where
the complete type IIB supergravity action has been build, and latter
in the independent study of \cite{Eric+05=IIB}.

A candidate bosonic action for Q7-branes  was considered in
\cite{Eric+06aD7} (under the name of `new seven--branes') and in
\cite{Dima+Eric07}. Although some issues of supersymmetry have been
discussed in \cite{Eric+06sD7} and in \cite{Dima+Eric07}, neither
the complete supersymmetric worldvolume action nor the complete
supersymmetric set of equations of motion for the Q7--branes were
proposed yet.

In this paper we explore the possibility to use the superembedding
approach
\cite{bpstv,hs96,hs2,hsw97,bst97,HS+Chu=PLB98,ABKZ,HSW+Chu=PLB98,Dima99,B00,Howe+Sezgin04,DpSL2}
(see \cite{Dima99,B00} for more references) to obtain a manifestly
supersymmetric description of Q7-brane dynamics. The superembedding
approach has already shown its usefulness for this type of problems.
For instance, the equations of motion for the M-theory 5-brane
(M5-brane) were obtained in \cite{hs2}  several months before the
covariant action was constructed in \cite{blnpst} an, independently,
in \cite{schw5}.

Furthermore, even if one is oriented on the covariant action, this
also could be constructed on the basis of superembedding approach.
We do not mean  complete superfield  action, like the so--called STV
action for lower dimensional Brink-Schwarz superparticles
\cite{stv}\footnote{STV is for Sorokin, Tkach and Volkov, the
authors of the pioneer paper \cite{stv}. (See also \cite{vz,stvz}
for related studies). Such actions are known for superparticles and
superstrings in superspaces with up to 16 supersymmetries, including
the heterotic string without heterotic fermions \cite{DGHS93},  as
well as for lower dimensional supermembranes (see
\cite{PBSG,Dima+2002} and \cite{Dima99} for review and further
references), but are not known neither for  D=10 type II superbranes
nor for D=11 M--branes. The problem appears also for heterotic
string, on the stage  when one tries to include heterotic fermions
(or heterotic bosons). A number of approaches to superfield
description of heterotic fermions were proposed
\cite{HeteroticF,Howe:1994ej}, but the most successful of them
\cite{Howe:1994ej} is restricted to the case of $SO(4)$ group,
rather than $SO(32)$ or  $E_8\otimes E_8$ charactersitic for the
anomaly free heterotic string theory.}. We rather refer on the way
of restoring the Green--Schwarz type action from the superembedding
approach which was  proposed in \cite{Howe-GAP} (and can be treated
as a bottom-up version of the generalized action principle for
superbranes \cite{bsv}\footnote{See \cite{Tonin91} for an earlier
approach to the action for heterotic string similar to the ones in
\cite{bsv,Howe-GAP}.}).

The analysis of possible closed 9--forms on type IIB target
superspace (which we carry out in Sec. 2 using the results of
\cite{DpSL2}) shows the existence of a candidate Wess--Zumino (WZ)
term, provided one assumes the presence of two linearly independent
gauge fields on the worldvolume\footnote{The same conclusion, and
also an expression for the (bosonic) Wess--Zumino  term in terms of
gauge potentials and pull--backs of supergravity fields can be found
in \cite{Dima+Eric07}.}. Under independent gauge fields we
understand such two worldvolume gauge potentials  that their
generalized field strengths involve different, linearly independent
combinations of the pull-backs of the NS-NS (Nevieu-Schwarz --
Nevieu-Schwarz) and RR (Ramond--Ramond) 2-form gauge potentials of
the type IIB supergravity\footnote{In other words, two independent
worldvolume gauge fields are Goldstone fields for different,
linearly independent combinations of the NS-NS and RR gauge
transformations.}. This WZ terms could also be used to construct the
complete action on the line of \cite{Howe-GAP} {\it provided there
exists its complete worldvolume superspace extension}. This implies
an existence of two linearly independent gauge supermultiplets in
the 7-brane worldvolume superspace $W^{(8|16)}$ with 8-bosonic and
16 fermionic `directions'.

Thus the search for a description of Q7--brane dynamics implies the
study of the embedding of the 7-brane worldvolume superspace
$W^{(8|16)}$ into the tangent superspace of type IIB supergravity
${\cal M}^{(10|16+16)}$, the simplest representative of which is the
flat type IIB superspace $\Sigma^{(10|16+16)}$, and to find a
possibility to describe supermultiplets with two independent
worldvolume gauge field potentials. (We turn to this study in Sec
3).

With quite general assumption, including the standard form of the
superembedding equation \cite{stv,bpstv,Dima99} and the gauge field
constraints motivated by the consistency of the linear
approximation, our results are negative: {\it there is no place for
Q7-brane in the superembedding approach} (see Sec. 4). In the light
of universality of the supermebedding approach, this might be
considered as an indication of {\it that Q7-brane is not a dynamical
brane but rather a supersymmetric ground state of the system of two
'standard' 7-branes} (we call these {\it SD7-branes}, also the name of
'$(p,q)$' seven branes can be used), each related to the {\it D7-brane}
(Dirichlet super-7--brane) by different SL(2)
transformation\footnote{In quantum theory $SL(2,\mathbb{R})$
symmetry characteristic of type IIB supergravity  is broken by Dirac
quantization conditions for the brane charges down to its
$SL(2,\mathbb{Z})$ subgroup. In this paper we find convenient,
following \cite{DpSL2}, to use the shortened notation SL(2).}.

Then, if supersymmetry is characteristic for the ground state only
and the excited states of the system are not supersymmetric, the
possible effective action for the two-brane system and the system of
interacting equations of motion would not be $\kappa$--symmetric,
while the ground state solution of the coupled equations of motion
would possess supersymmetry. Such a picture was observed in
\cite{BK99} in an attempt to develop complete supersymmetric
Lagrangian description for the interacting
superstring---super-D$p$-brane dynamical system. The conjecture
suggested by our study is that the bosonic Q7--brane actions
considered in \cite{Eric+06aD7,Dima+Eric07} are effective actions of
this type, which do not allow a supersymmetric and
$\kappa$--symmetric completion but do allow for a supersymmetric
ground state solution.

In concluding Sec. \ref{Conclusion} we discuss our results and  the
arguments in favor of the above conjecture as well as their possible
significance for understanding the general aspects of multi-brane
interactions and, in particular, the old-standing problem of
constructing the supersymmetric and Lorentz SO(1,9) symmetric
D$p$-brane action\footnote{\label{footnote5}This can be reformulated
as a problem of lack of supersymmetry and Lorentz symmetry in the
Myers action \cite{Myers:1999ps}. Notice that some progress in this
direction was reached for the cases of low dimensions $D$, low
dimensional branes  and low co-dimensional branes
\cite{Dima01,Drummond:2002kg,Dima+Panda=03}. Also a very interesting
{\it minus one quantization} approach using {\it string with
boundary fermions} was proposed in \cite{Howe+Linstrom+Linus}; the
quantization of the boundary fermions should reproduce the Myers
action in this scheme (hence 'minus one quantization' name above).
An attempt to reformulate the matrix diffeomorphism invariance as
base-point-independence was discussed in \cite{Matrixdiff} for
bosonic D--branes.}.

To make the text lighter, we moved technical details to Appendices
A-E. Our notation and conventions are described in Appendices A and
B. We denote the bosonic and fermionic supervielbein forms of
general type IIB supergravity superspace by
\begin{eqnarray}\label{Eua-cE}
 {E}^{{\underline{A}}}= ({E}^{{\underline{a}}}, {\cal E}^{{\underline{\alpha}}})\; , \qquad
 {\cal E}^{{\underline{\alpha}}}:= {\cal E}^{{{\alpha}i}}:= (E^{\alpha 1}\, , \; E^{\alpha 2}) \;
 ,  \qquad
\end{eqnarray}
$\underline{a}, \underline{b}, \underline{c}=0,1,\ldots, 9$ are
tangent space vector indices, ${{\alpha}}, {{\beta}},
{{\gamma}}=1,\ldots , 16$ are $D=10$ Majorana-Weyl spinor indices.
The supervielbein obeys the type IIB supergravity constraints
\cite{HoweWest83}, the most important of which is
$T_{{\underline{\alpha}}{\underline{\beta}}}{}^{\underline{a}}=
-2i\underline{\sigma}^{\underline{a}}_{{\underline{\alpha}}{\underline{\beta}}}
:= -2i\sigma^{\underline{a}}_{\alpha\beta} \delta_{ij}$, where
$\sigma^{\underline{a}}_{\alpha\beta}$ are real symmetric $16\times
16$ matrices (generalized Pauli matrices or Klebsh-Gordan
coefficients for $SO(1,9)$).  Together with conventional constraints
and their consequences this results in the following form of the
type IIB superspace bosonic  torsion 2-form, which can be loosely
called {\it bosonic torsion constraint},
\begin{eqnarray}
\label{Tua=} {T}^{\underline{a}} := D{E}^{\underline{a}}=  -i {\cal
E}\wedge \underline{\sigma}^{\underline{a}}{\cal E}\, \; , \qquad
\underline{\sigma }^{\underline{a}}:= \underline{\sigma
}^{\underline{a}}_{{\underline{\alpha}}{\underline{\beta}}} :=
\sigma^{\underline{a}}_{\alpha\beta} \delta_{ij}\; . \quad
\end{eqnarray}
%Other constraints of type IIB supergravity and their consequences
%useful in the present study are collected in the Appendix B.

\section{
Problem statement. Candidate Wess--Zumino term for a hypothetical
Q7-brane requires two worldvolume gauge fields. } \label{Ch-E=Co}

Wess--Zumino (WZ) term describes coupling of a brane to
antisymmetric tensor fields of the supergravity multiplets, as well
as to axion and dilation (for D$7$-branes only axion contributes to
the WZ term).

The candidate WZ term for Q7--brane was obtained  in
\cite{Dima+Eric07} from the requirement of gauge invariance.
Although the  bosonic construction of \cite{Dima+Eric07} allows for
a straightforward supersymmetric generalization, we will present
here its independent derivation, based on the results of
\cite{DpSL2}. This will allow us to see the relation between the
D7-brane Wess--Zumino term and a triplet of nine forms which gives
rise to the candidate WZ term for Q7, and will also demonstrate the strong
necessity of the introduction of the second gauge field to write a
closed covariant nine-form different from the D7--brane WZ term.

The coupling to the type IIB scalars (axion and dilaton) was studied
in \cite{Eric+06aD7,Dima+Eric07} and the properties of such coupling
were used intensively in the studies of these papers. However, if
Q7-brane existed as a dynamical object, its action would make sense
in any superspace background obeying the on-shell type IIB
supergravity constraints (as it is for the case of the standard
D7-brane action of \cite{Dpac}). In particular, it should exist the
formulation in flat type IIB superspace, with vanishing axion and
dilation.

It is natural to begin the study by considering this simplest case.
The flat type IIB superspace $\Sigma^{(10|16+16)}$  can be
characterized by the following superspace constraints (which differs
from (\ref{Tua=}) by absence of the spin connection;
$DE^{\underline{a}}=dE^{\underline{a}}- E^{\underline{b}}\wedge
w_{\underline{b}}{}^{{\underline{a}}}$ in (\ref{Tua=}))
\begin{eqnarray}
\label{Tub=flat}  d{E}^{\underline{a}}= -i (E^{{\alpha}1} \wedge
E^{{\beta}1} - E^{{\alpha}2} \wedge E^{{\beta}2})
\sigma^{\underline{a}}_{{\alpha}{\beta}}= -i {\cal
E}\wedge \underline{\sigma}^{\underline{a}} {\cal E} \; , \qquad \\
\label{Tf=flat} d{\cal E}^{\underline{\alpha}}=0\; \qquad
\Leftrightarrow \qquad \cases{ dE^{{\alpha}1}=0\; , \cr
dE^{{\alpha}2}=0\; . \qquad } \qquad
\end{eqnarray}
This can be solved in terms of local coordinates by identifying
supervielbein with the D=10 type IIB counterparts of the
Volkov-Akulov supersymmetric 1-forms \cite{VA72},
\begin{eqnarray}
\label{Eub=flat} {E}^{\underline{a}}= d{X}^{\underline{a}}-i
d\theta^1 \sigma^{\underline{a}}\theta^1-i d\theta^2
\sigma^{\underline{a}}\theta^2= d{X}^{\underline{a}}-i
d\Theta\underline{\sigma}^{\underline{a}}\Theta\; , \qquad \nonumber \\
{\cal E}^{\underline{\alpha}}=d\Theta^{\underline{\alpha}}\;  \qquad
\Leftrightarrow \qquad \cases{E^{{\alpha}1}=d\theta^{{\alpha}1} \; ,
\cr E^{{\alpha}2}=d\theta^{{\alpha}2}\; .} \qquad
\end{eqnarray}

\subsection{Closed 9-forms in type IIB superspace}

The Wess--Zumino term for a 7--brane in flat superspace  can be
characterized by a closed invariant 9--form (this can be identified
with 9--cocycle of the Chevalley-Eilenberg cohomology
\cite{JdA+PKT89,JdA2000}). To find all the possible candidate 9-form
WZ terms, one has to carry out the analysis of all the lower order
invariant forms the wedge products of which can be used in the
construction. Fortunately, we do not need to perform calculation
ourselves as all such forms appear as flat superspace limit of the
superspace field strengths  of the $SL(2)$ covariant formulation of
the superspace type IIB supergravity elaborated in \cite{DpSL2}. In
addition to derivatives of axion and dilaton superfields, which are
set to zero in the flat superspace limit ($F_R^{(1)}=0$), the list
of the field strength of SL(2) invariant formulation of type IIB
supergravity \cite{DpSL2} includes the doublet of three--forms,
$F_R^{(3)}$, $R=1,2$, the singlet 5-form $F^{(5)}$, the doublet of
seven--forms, $F_R^{(3)}$, and the triplet of the nine--forms
$F_{RS}^{(9)}=F_{(RS)}^{(9)}$\footnote{In this section we mainly use
the SL(2) covariant notations of \cite{DpSL2} ($SL(2)$ symmetry
reduces to $SO(2)$ when axion and dilaton are set to zero, which is
the case in flat superspace); below we give the relation to the
NS-NS and R-R fields in a more familiar 'D-brane basis', which is
not SL(2) covariant.}. These obey the following Bianchi identities
\cite{DpSL2}
\begin{eqnarray}
\label{BIs=flatSL} dF_R^{(3)}=0\; , \qquad dF^{(5)}=-\epsilon^{RS}F_R^{(3)}\wedge
F_S^{(3)}\; , \qquad \nonumber \\  dF_R^{(7)}=F_R^{(3)}\wedge F^{(5)}\; , \qquad
dF_{RS}^{(9)}=F_{(R}^{(3)}\wedge F_{S)}^{(7)}\; , \qquad
\end{eqnarray}
In flat type IIB superspace these field strengths are represented by
the following supersymmetry invariant forms
\begin{eqnarray}
\label{F3R=flat} F_R^{(3)} &=& -i{\cal E}^{{\alpha}i} \wedge {\cal
E}^{{\beta}j}\wedge \bar{\sigma}^{(1)}_{{\alpha}{\beta}}\,
(\tau_R)_{ij} \; , \qquad \\ \label{F5=flat} F^{(5)} &=& \; i {\cal
E}^{{\alpha}i} \wedge {\cal E}^{{\beta}j}\wedge
\bar{\sigma}^{(3)}_{{\alpha}{\beta}}\, \epsilon_{ij} \; , \qquad
\\ \label{F7R=flat} F_R^{(7)} &=& \; i {\cal E}^{{\alpha}i} \wedge
{\cal E}^{{\beta}j}\wedge \bar{\sigma}^{(5)}_{{\alpha}{\beta}}\,
(\epsilon_{RS}\tau^S)_{ij} \; , \qquad
\\ \label{F9RS=flat} F_{RS}^{(9)}\! &=& \; {i\over 2}\; {\cal E}^{{\alpha}i} \wedge
{\cal E}^{{\beta}j}\wedge \bar{\sigma}^{(7)}_{{\alpha}{\beta}}\,
\epsilon_{ij} \delta_{RS} \; , \qquad
\end{eqnarray}
where
\begin{eqnarray}
\label{barsig=} \bar{\sigma}^{(2n+1)}_{{\alpha}{\beta}}  & :=
{1\over (2n+1)!} E^{a_{2n+1}}\wedge ... \wedge E^{a_{1}}\;
\sigma_{a_{1}... a_{2n+1}}{}_{{\alpha}{\beta}}\; , \qquad
\end{eqnarray} and in flat
superspace
 $(\tau_R)_{ij}=\delta_R^r (\tau_r)_{ij}$ with
\begin{eqnarray}
\label{tauR}
(\tau_r)_{ij}={1\over \sqrt{2}}\left(\sigma_3\, , \, \sigma_1
\right)
\; . \qquad
\end{eqnarray}
Clearly, these matrices are invariant under $SO(2)$ but not under
the $SL(2)$ group; in general curved supergravity background
$(\tau_R)_{ij}={\cal U}_R{}^r (\tau_r)_{ij}$, where ${\cal U}_R{}^r$
is the axion-dilatom matrix providing the bridge between $SO(2)$ and
$SL(2)$ groups.

This is the place to notice that the original papers of D-brane
actions \cite{Dpac} used the NS-NS and RR field strengths $H_3$,
$H_7$ and $R_{2n+1}$ which obey the Bianchi identities (we set
$R_1=0$, as it is in flat superspace)
\begin{eqnarray}
\label{BI=dH} dH_3=0 \; , \qquad dH_7= - R_3\wedge R_5\; , \qquad
dR_{2n+1}=-H_3\wedge R_{2n-1} \; . \qquad
\end{eqnarray}
Comparing Eqs. (\ref{F3R=flat})--(\ref{F9RS=flat})  with
(\ref{BI=dH}) one finds that the $SL(2)$ multiplets of field
strength can be expressed  as
\begin{eqnarray}
\label{F3=(H,R)} F^{(3)}_R= {1\over \sqrt{2}} \left(\matrix{H_3\cr -
R_3}\right) \; , && \qquad F^{(5)}=-R_5\; , \qquad \\
\label{F7=(H,R)}
 F^{(7)}_R= {1\over \sqrt{2}} \left(\matrix{R_7\cr - H_7}\right) \; , &&
\qquad F^{(9)}_{RS}=- {1\over 2}\delta_{RS}R_9\; . \qquad
\end{eqnarray}
Notice that in our notation for the forms in  'D-brane basis', which
we use in the main text, the number in subscript gives the order of
form, while in the SL(2) covariant formalism, which we use only in
this section, we keep the original notation of \cite{DpSL2} in which
the order of form is given in superscript by the number in brackets.

Now, to search for the candidate WZ term we need to use only Bianchi
identities and Eqs. (\ref{F3=(H,R)}), (\ref{F7=(H,R)}). The explicit form of
(\ref{F3R=flat})--(\ref{F9RS=flat}) is needed for calculating
fermionic variations of the candidate action,  but we will not turn
to this issue in this paper.

Using the above field strengths one cannot construct any closed and
supersymmetric invariant 9-form. This corresponds to non-existence
of a super-7-brane without worldvolume gauge fields in type IIB
superstring theory.

To write the WZ term of the D7--brane \cite{Dpac} one introduces the
field strength ${\cal F}_2=dA-\hat{B}_2$ of the worldvolume gauge
field $A=d\xi^mA_m(\xi)$. These contains pull--back to the
worldvolume $\hat{B}_2$ of the NS--NS two form gauge potential
$B_2$, the field strength of which is $H_3=dB_2$. Below in this
section we will not distinguish pull-back from the form notationally
i.e. we will omit hat symbols when this would not lead to a possible
confusion. Thus
\begin{eqnarray}
\label{F2=dA-B2} {\cal F}_2=dA-B_2\; ,\qquad d{\cal F}_2=-H_3 \; .
\qquad
\end{eqnarray}

Completing the set of (pull--backs of) RR and NS-NS forms by the
worldvolume field strength (\ref{F2=dA-B2}), one can find a closed
invariant 9--form. It gives the (formal exterior derivative of the)
D$7$ brane WZ term \cite{Dpac} and reads\footnote{We use the
notation $d{\cal L}_8^{WZ-D7}$ as far as the nine form
(\ref{LWZD7=}) is exact in de Rahm cohomology, but call it closed
because it represents a nontrivial cocycle of the
Chevalley--Eilenberg cohomology \cite{JdA+PKT89}. This is to say,
the form ${\cal L}_8^{WZ-D7}$ does exist (hence, exactness in de
Rahm cohomology), but it is not invariant under supersymmetry (hence
$d{\cal L}_8^{WZ-D7}$ is not exact in the Chevalley--Eilenberg
cohomology). We refer on \cite{JdA2000} for more discussion and
intriguing applications. }
\begin{eqnarray}
\label{LWZD7=} d{\cal L}_8^{WZ-D7}=R_9+ {\cal F}_2 \wedge R_7+
{1\over 2}{\cal F}_2 \wedge {\cal F}_2 \wedge R_5+ {1\over 3!}{\cal
F}_2 \wedge {\cal F}_2 \wedge {\cal F}_2 \wedge R_3 \; . \qquad
\end{eqnarray}
This covariant 9--form is unique, as far as only one gauge field with the generalized
field strength related to the NS-NS two form potential
(\ref{F2=dA-B2}) is introduced.

To search for 9--form describing a candidate WZ term of the
hypothetical Q7-brane action, first one has to introduce a doublet
of the worldvolume field strengths involving two worldvolume gauge
potentials and pull--backs of the doublet of two-form gauge
field potentials (NS-NS and RR potentials),
\begin{eqnarray}
\label{F2R=(H,R)} {\cal F}^{(2)}_R:= dA^{(1)}_R - C_R^{(2)}= {1\over
\sqrt{2}} \left(\matrix{\; {\cal F}_2\cr - \tilde{{\cal
F}}_2}\right)={1\over \sqrt{2}} \left(\matrix{\; dA-B_2 \cr -
(d\tilde{A}- C_2)}\right) \; , \qquad
\end{eqnarray}
so that
\begin{eqnarray}
\label{dF2R=(H,R)} d{\cal F}^{(2)}_R:= F_R^{(3)}\qquad
\Leftrightarrow \qquad  \left\{\matrix{\; H_3=- d{\cal F}_2 \cr
R_3=- d\tilde{{\cal F}}_2}\right. \; . \qquad
\end{eqnarray}

Now, starting from the triplet of the nine-form field strength,
$F_{RS}^{(9)}$ (\ref{F9RS=flat})\footnote{One easily notices that
this triplet of 9-forms over flat superspace has actually only one
independent component. Although in the case of curved superspace the
other two nonvanishing contributions appear (the purely bosonic
form contributions are related to derivatives of axion and dilaton
by dualities, and also $\delta_{ij}$ in (\ref{F9RS=flat}) is
replaced by a bilinear of the axion--dilaton matrix, see
\cite{DpSL2}) the fact of reduction of a triplet of 9-form
$F^{(9)}_{RS}$ to a singlet in the flat superspace limit already
should rise suspicions concerning the existence of Q7-branes as
dynamical objects.}, and searching for an SL(2) covariant closed
form $d{\cal L}^{(8)}_{RS}$, one finds the following triplet of
closed 9--forms
\begin{eqnarray}
\label{LWZRS=} d{\cal L}^{(8)}_{RS}= {F}^{(9)}_{RS}-
 {\cal F}^{(2)}_R \wedge {F}^{(7)}_{S} +
{1\over 2} {\cal F}^{(2)}_R  \wedge  {\cal F}^{(2)}_S  \wedge F_5+ {1\over 4} {\cal
F}^{(2)}_R \wedge  {\cal F}^{(2)}_S  \wedge \epsilon^{R'S'}\,  {\cal F}^{(2)}_{R'}
 \wedge  {\cal F}^{(2)}_{S'}  \; . \qquad
\end{eqnarray}

Notice that $(1,1)$ component of the triplet is similar, but not
identical to the D$7$-brane WZ term (\ref{LWZD7=}). Namely
\begin{eqnarray}
\label{LWZRS=LWZD7+} d{\cal L}^{(8)}_{R=1\, , \, S= 1}= - {1\over 2}
d{\cal L}^{WZ-D7}_{8} +  {1\over 48}({\cal F}_2  \wedge {\cal F}_2
\wedge {\cal F}_2  \wedge R_3 + 3 {\cal F}_2  \wedge{\cal F}_2
\wedge\tilde{{\cal F}}_2  \wedge H_3)\; .
\end{eqnarray}
The difference is one of the representatives of the family of 
exact  forms which appear in the presence of two worldvolume gauge
fields,
\begin{eqnarray}
\label{calCqp=} {\cal C}^{(q,p)}_{2q+2p+5} &=& (q+1){\cal F}^{\wedge
q}_2 \wedge\tilde{{\cal F}}{}^{\wedge (p+1)}_2\wedge H_3 + (p+1)
{\cal F}^{\wedge (q+1)}_2 \wedge\tilde{{\cal F}}{}^{\wedge
p}_2\wedge R_3\; = \qquad \nonumber \\ &=& - d \left({\cal
F}^{\wedge (q+1)}_2 \wedge\tilde{{\cal F}}{}^{\wedge (p+1)}_2
\right)\,  . \qquad
\end{eqnarray}
Here ${\cal F}^{\wedge q}_2= \underbrace{{\cal F}_2 \wedge \ldots
\wedge {\cal F}_2}_q$, {\it etc.}

\subsection{SD7--branes, Q7-branes and multiple (S)D-brane systems}

A typical representative of the family of hypothetical Q7-branes
would be characterized by the Wess-Zumino term \begin{eqnarray}
\label{LWZQ7=QL} d{\cal L}^{(8) WZ-Q7}= Q^{RS}d{\cal L}^{(8)}_{RS}\,
, \qquad
\end{eqnarray} with a constant symmetric
$2\times 2$ matrix $Q^{RS}$ the components of which give us,
generically, three integer charges characterizing this 7-brane.
Generic case corresponds to non-degenerate matrix $Q^{RS}$,
$det(Q^{RS})\not=0$. According to \cite{Dima+Eric07}, the cases of
charges forming the Q-matrix with positive determinant correspond to
the Q7-branes while matrices with negative determinants do not
correspond to any brane\footnote{Notice that here we prefer to deal
with the symmetric Q-matrix, which in notation of \cite{Dima+Eric07}
reads $Q=\left(\matrix{p & -r/2 \cr  -r/2 & q}\right)$. Its relation
with the traceless matrix $\mathbb{Q}=\left(\matrix{r/2 & p \cr -q &
-r/2}\right)$, mainly used in \cite{Dima+Eric07} to describe
Q7--brane charges, is given by $Q=\mathbb{Q}\, i\tau_2$. Referring
on three independent charges,  Q7-branes are also called
$(p,q,r)$-brane, and the condition $det Q= det \mathbb{Q}>0$ implies
$pq>{r^2\over 4}$.} (we will comment on this latter).

When determinant of the charge matrix is zero, its rank is 1 (if
nonvanishing) and, hence, it can be expressed in term of one SL(2)
vector, $Q^{RS}=q^R\, q^S$. The D7-brane correspond to
$q^R=\left(\matrix{1 \cr 0}\right)$ and
$Q^{RS}=\delta^{R}_1\delta^{S}_1$. The general $Q^{RS}=q^R\, q^S$
with $q^R\not=(0,0)$ corresponds to branes related to the D7-brane
by SL(2) transformations. We call these {\it SD7-branes}.

We should stress that, in the light of the $SL(2)$ duality
invariance of the type IIB theory, the choice of D7-brane among
SD7-branes  is purely conventional. The SL(2) covariant description
of the SD7-branes has been constructed in \cite{DpSL2}. This makes
clear that the property to have Wess--Zumino term expressed through
only one gauge field cannot be specific just for one D7-brane.
Indeed, the same expression with ${\cal F}_2$ replaced by
$\sqrt{2}\, q^R{\cal F}^{(2)}_R$ would serve for the Wess--Zumino
term of the SD7-brane with the charge matrix $Q^{RS}=q^R\, q^S$.

To resume, the charge matrices of different $7$--branes are
described by the table
\begin{eqnarray}\label{D7-Q7}
\matrix{D7-brane \quad & \leftrightarrow & \quad
Q^{RS}=\delta^{R}_1\delta^{S}_1\cr SD7-brane \quad & \leftrightarrow
& \quad Q^{RS}=q^R\, q^S \cr Q7-brane \quad & \leftrightarrow &
\quad det(Q^{RS})> 0  }
\end{eqnarray}
The last line actually can be equivalently written in the form of
\begin{eqnarray}\label{Q7=q+q}
Q7-brane \quad & \leftrightarrow \quad Q^{RS}=\pm (q_1^R\, q_1^S +
q_2^R\, q_2^S ) \qquad
\end{eqnarray}
with nonvanishing doublets of charges $q_1^R$ and $q_2^R$. This
suggests to consider {\it Q7-brane as  a bound state of two
SD7--branes} characterized by charges $q_1^R$ and $q_2^R$,
respectively. More precisely, this treatment corresponds to sign
plus in (\ref{Q7=q+q}), while in the case of minus sign one should
rather speak about bound state of two anti-SD7-branes. Notes that
the case with different signs for the first and second
contributions, which would correspond to a system of SD7-brane and
an anti-SD7-brane, are excluded by the conditions of having $det
Q>0$ \cite{Dima+Eric07}.

Thus, when considering Q7 as an interacting system of two
SD$7$-branes, the seemingly mysterious requirement of $det Q>0$ just
corresponds to the well known fact that the supersymmetry is broken
in the system including a brane and an anti-brane.

The possibility to be treated as a system composed  of two
SD7--branes does not prevent Q7-brane from being described by an
effective action and, in this sense, from being a dynamical object.
However, in this case one should expect, at least, that such a
hypothetical effective action for a bound state of two SD7-branes
does possess certain symmetries, including Lorentz invariance and
supersymmetry. If this is not the case, in particular, if an
effective action does not possesses supersymmetry, one may think
that the Q7--brane solution \cite{Eric+06aD7} does  not correspond
to a dynamical brane but rather to  a particular {\it ground} state
of a system of two interacting SD7--branes which does possesses
supersymmetry (in distinction to a generic states of this system).
An analogy comes from the study of supergravity solutions describing
intersecting plane branes \cite{P+PKT96,kappaInt}; in general such
solutions are not supersymmetric, but supersymmetry appears for
certain angles (the complete supersymmetry characteristic for one
brane in the case of coincidence, one half of the supersymmetry of
the one-brane solution for the orthogonal intersection, {\it etc.}).
We will be coming back to this point in the next sections and,
particularly, in the concluding Sec. \ref{Conclusion}.

To conclude this section, let us notice a similarity of the problem
of searching for the action for Q7--brane with the problem of
Lagrangian description of the multiple--D$p$--brane systems, which
becomes transparent after understanding the possible treatment of
Q7-brane as a coupled system of two SD7--branes. The problem with
Q7--brane action, which looks relatively simple in the light of do
not expecting a non-Abelian structure, might hence provide new
insights for the multiple--D$p$--brane system.\footnote{Let us
recall that the commonly used Myers action \cite{Myers:1999ps},
which predicted the 'dielectric brane effect' and was obtained by
using T-duality arguments, do not possess neither supersymmetry nor
Lorentz symmetry; see also footnote \ref{footnote5}. }

\subsection{Superembedding--based method to search for hypothetical
Q7-brane action. Problem statement. \label{2.3}}

The action for SD7-brane was constructed in \cite{DpSL2}  by the
method proposed in \cite{Howe-GAP} starting from the superembedding
description \cite{hs96,bst97}. Roughly, this procedure can be split
on the following stages. One i) finds the Wess--Zumino term, ii)
lifts it to the maximal worldvolume superspace of the $p$--brane,
{\it restricted by the superembedding equation} (see below) and then
iii) uses this superspace form to search for the kinetic,
(presumably) Dirac--Born-Infeld-like (DBI) part of the action in an
algorithmic manner. The detailed description of the stage (iii) is
not needed for our purposes here, it can be found in 
\cite{Howe-GAP,DpSL2}. The first stage (i) was the subject of our
Secs. 2.1 and 2.2. The main subject of our study below will be the
possibility to progress in the second stage (ii).

Of course, the DBI part may be also constructed by searching for the
$\kappa$--symmetric completion of the Wess--Zumino term, but the
superembedding approach based method of \cite{Howe-GAP,DpSL2} is
more algorithmic and, hence, more conclusive in the case of negative
result (which we will actually arrive at in the case of Q7-branes).

Now, having in hand the candidate Wess--Zumino term for the
Q7-brane, it is natural to apply this superembedding-based method in
our search for a complete Q7-brane action. The first stage in this
direction should be, as we commented above, to lift the candidate
Wess--Zumino term to the complete worldvolume superspace (N=1, d=8
superspace in the case of type IIB 7-branes) subject to the
superembedding equation (discussed below). However, {\it this
inevitably implies that two gauge fields living on the hypothetical
Q7-brane worldvolume, are lifted to the worldvolume superspace}.

Thus, the first question to ask is whether a 7-brane worldvolume
superspace can carry two  super-1-form gauge potentials which are
essentially different in the sense of that their invariant field
strengths contain pull--backs of the different (linear independent)
combinations of the NS-NS and R-R two-form potentials.

This will be the main subject of Sec. 4. But before turning to it,
in the next Sec. 3 we describe the general features of the
superembedding approach, specifying it for type IIB 7-branes,  and
its application to obtain D$7$--brane equations of motion.

\bigskip

\section{Superembedding approach to type IIB 7-branes and
description of D7-brane dynamics}

\label{SuperEmb}

\subsection{Superembedding equation for type IIB 7-branes}

\subsubsection{Worldvolume superspace, pull--backs of target superspace
superforms and superembedding equation}

To write the most general and universal form of the {\it
superembedding equation} for a super--$p$--brane in D=10 type IIB
supergravity background, let us first denote the $d=p+1\leq 10$
local bosonic coordinates and $16$ fermionic coordinates of the
worldvolume superspace $W^{(p+1|16)}$ by $\zeta^{{\cal M}}=(\xi^m ,
 \eta^{\check{q}} )$. Then let us notice that the embedding of
$W^{(p+1|16)}$ into the tangent type IIB superspace
$\Sigma^{(10|32)}$ with coordinates $Z^{\underline{M}}=
(x^{\underline{m}}\; , \; \theta^{\check{\alpha}1}\; , \;
\theta^{\check{\alpha}2}\;)$ can be described parametrically by
specifying the set of coordinate functions, the {\it worldvolume
superfields} $\hat{Z}^{\underline{M}}(\zeta)$,
\begin{eqnarray}
\label{WinS}  W^{(p+1|16)}\in \Sigma^{(10|32)} &:& \qquad
Z^{\underline{M}}= \hat{Z}^{\underline{M}}(\zeta)\; ,  \qquad
\nonumber
\\  && \zeta^{{\cal M}}=(\xi^m \; , \; \eta^{\check{q}} )\; , \qquad  \hat{Z}^{\underline{M}}(\zeta)  = (\hat{x}^{\underline{m}}(\zeta)\; ,
\hat{\theta}^{\check{\alpha}1}(\zeta )\; , \;
\hat{\theta}^{\check{\alpha}2}(\zeta ))\; , \qquad
\end{eqnarray}
with $ {\underline{m}}= 0,1,\ldots , 9 \;$, $\check{\alpha}=1,\ldots
16\;$,
  $\check{q}=1,\ldots 16\;$ and $m=0,1,\ldots 7\;$
in the case of type IIB 7-branes. The superembedding equation is
imposed on these coordinate functions.

Denoting the supervielbein of the worldvolume superspace
$W^{(8|16)}$ by
\begin{eqnarray}
\label{eA=ea+} e^A= d\zeta^{{\cal M}} e_{{\cal M}}{}^{A}(\zeta) =
(e^a\; , \; e^q) \; , \qquad a=0,1,\ldots , 7\; , \qquad q=1,\ldots,
16 \; ,
\end{eqnarray}
one can decompose the pull--back
$\hat{E}^{\underline{A}}:=E^{\underline{A}}(\hat{Z})$ of the
supervielbein of the target type IIB superspace
$E^{\underline{A}}=dZ^{\underline{M}}
E_{\underline{M}}{}^{\underline{A}}(Z)=(E^{\underline{a}}\, ,
E^{\underline{\alpha}1}, E^{\underline{\alpha}2})$, Eq.
(\ref{Eua-cE}), on the basis (\ref{eA=ea+}). In general, such a
decomposition reads
\begin{eqnarray}
\label{hEa=b+f}
 \hat{E}^{\underline{A}}:= E^{\underline{A}}(\hat{Z})=
d\hat{Z}^{\underline{M}}
E_{\underline{M}}{}^{\underline{A}}(\hat{Z}) = e^b \hat{E}_b^{\,
\underline{A}} + e^q \hat{E}_q^{\, \underline{A}} \; , \qquad
\end{eqnarray}
where $\hat{E}_b^{\, \underline{A}}:= e_b^{{\cal M}} \partial_{{\cal
M}} \hat{Z}^{\underline{M}}
E_{\underline{M}}{}^{\underline{A}}(\hat{Z})$ and $\hat{E}_q^{\,
\underline{A}}:= e_q^{{\cal M}}\partial_{{\cal M}}
\hat{Z}^{\underline{M}}
E_{\underline{M}}{}^{\underline{A}}(\hat{Z})$.

The superembedding equation states that the fermionic component of
the pull--back of the bosonic supervielbein form vanishes,
\begin{eqnarray}
\label{SembEq}
 \fbox{$\hat{E}_q{}^{\underline{a}}:= \nabla_q
 \hat{Z}^{\underline{M}}\,
E_{\underline{M}}{}^{\underline{a}}(\hat{Z}) =0 \;$}\;  ,  \qquad
\nabla_q:=e_q^{{\cal M}}(\zeta) \partial_{{\cal M}}\, , \quad
\zeta^{{\cal M}}=(\xi^m  ,  \eta^{\check{q}} )\; .
\end{eqnarray}
This superembedding equation was first obtained form the STV action
for $D=3,4$ dimensional superparticle \cite{stv} and was used as a
basis to develop superembedding approach for D=10 superstrings and
D=11 supermembrane (M2-brane) in \cite{bpstv} (see \cite{Dima99} for
more references). The superembedding equation for Dp-branes and
M5-brane were used to derive their equations of motion, respectively in
\cite{hs96} and \cite{hs2}, before the complete covariant action was
found, respectively in \cite{Dpac} and \cite{bpstv,schw5}.

\bigskip

\subsubsection{Linearized superembedding equation in 'static gauge'
and Goldstone superfields} \label{3.2.1}

To create some feeling of the superembedding equation, it is useful
to consider its linearized version in flat target superspace (see
\cite{hs96} for the case of D$p$-branes). In this approximation a
$7$--brane is described by one complex (two real) bosonic
superfield(s) $\widetilde{X}^z(\zeta )
=(\widetilde{\bar{X}}{}^{\bar{z}}(\zeta ))^*$  and $16$ pseudo-real
fermionic superfields ${W}^{q}(\zeta
)=\gamma^0_{qp}({W}^{p}(\zeta))^*$ which obey
\begin{eqnarray}
\label{SEmbl=z} D^0_q \tilde{X}{}^z = -2i (\delta +
i\gamma^9)_{qp}{W}^p \; ,  \qquad D^0_q \tilde{\bar{X}}{}^{\bar{z}} =
-2i (\delta - i\gamma^9)_{qp}{W}^p \;
  \; . \qquad
\end{eqnarray}
Here $D^0_q={\partial\over \partial \eta^q} + 2i \theta^p \gamma^a_{pq}\partial_a$,  $\;\; p,q=1,\ldots, 16$ are pseudo-Majorana d=8 spinor indices, $\gamma^{a}_{pq}$
 are the $d=8$ gamma matrices, $a=0,1,...,7$, and $\gamma^9=\gamma^0\gamma^1\ldots \gamma^7$. See Appendix A for
further details.

To arrive at (\ref{SEmbl=z}) one uses the worldvolume diffeomorphism
symmetry to fix the so--called static gauge in which the coordinates
of the worldvolume superspace are identified with $8$ of $10$
bosonic and $16$ of $32$ fermionic coordinate functions,
\begin{eqnarray}
 \label{XTh=coord}
 \xi^{{a}}= \hat{X}{}^{a}(\xi ,\eta) -  i W(\xi , \eta)\gamma^a\eta \, , \qquad
\eta^q=  \hat{\theta}{}^{1q}(\xi ,\eta):=
\hat{\theta}{}^{1\alpha}(\xi ,\eta)\delta_\alpha{}^q  \; . \qquad
 \qquad
\end{eqnarray}
The remaining coordinate superfunctions, $\hat{X}{}^{8}$, $
\hat{X}{}^{9} $ and $\hat{\theta}{}^{2\alpha}$, are associated to
the Goldstone superfields $\widetilde{X}^z(\zeta )\equiv
\widetilde{X}^z(\xi , \eta )$  and ${W}^{q}(\zeta )\equiv
{W}^{q}(\xi , \eta )$. In linearized approximation it is convenient
to define these by
\begin{eqnarray}
 \label{XTh=fields}
 \widetilde{X}^z(\xi , \eta ):= \hat{X}{}^{8}+i \hat{X}{}^{9} + {i}W(\delta +
i\gamma^9)\eta  \, \quad and  \quad  W^q(\xi, \eta) :=-
\hat{\theta}{}^{2\alpha}\delta_\alpha{}^p\gamma^9_{pq} - \eta^q \;
 \qquad
\end{eqnarray}
(equivalent to $\theta^2= (\theta^1 +W)\gamma^9= (\eta
+W)\gamma^9$). This choice results in the simple form
(\ref{SEmbl=z}) of the linearized superembedding equation which
provides the superfield description of the $d=8$, ${\cal N}=1$
scalar supermultiplet.

\bigskip

\subsubsection{Equivalent form of superembedding equation,
induced worldvolume supervielbein and moving frame variables}

In the discussion below it is convenient to use the worldvolume
supervielbein induced by superembedding. This implies, in
particular, that the bosonic supervielbein form $e^a$ is expressed
by
\begin{eqnarray}\label{Eua=ea}
\hat{E}^a:= \hat{E}^{\underline{b}} u_{\underline{b}} ^{\; a} = e^a
\;   \qquad
\end{eqnarray}
in terms of the pull--back to $W^{(8|16)}$ of the bosonic type IIB
supervielbein, $\hat{E}^{\underline{b}}$, and the {\it moving frame
variables} $u_{\underline{b}} ^{\; a}$. These are eight ortogonal
and normalized vectors
\begin{eqnarray}\label{utut=1tt}u_{\underline{a}} ^{\; a}
\eta^{\underline{a}\underline{b}}u_{\underline{b}} ^{\; b}=\eta^{ab}
\;  , \qquad a,b=0,1,\ldots , 7\; , \qquad \underline{a}\, , \,
\underline{b}=0,1,\ldots , 9\; . \qquad
\end{eqnarray}
In flat superspace  ${E}^{\underline{b}}$ has the form of
(\ref{Eub=flat}); in general it obeys the type IIB supergravity
constraints, the most essential of which, $T_{\alpha i\; \beta
j}{}^{\underline{b}}= -2i
\delta_{ij}\sigma_{\alpha\beta}^{\underline{b}}$, is included in
(\ref{Tua=}).

One can complete the above set of orthogonal and normalized vectors
$u_{\underline{a}} ^{\; a}$ ($a=0,1,\ldots, 7)$ by two orthogonal to
them and also normalized vectors $u_{\underline{a}} ^{\; i}=
(u_{\underline{a}} ^{\; 8}, u_{\underline{a}} ^{\; 9})$, $\;
u_{\underline{a}} ^{\; i}u^{\underline{a}\; a}=0$,
$u_{\underline{a}} ^{\; i}u^{\underline{a}\; j}=-\delta^{ij}$. As
far as $u_{\underline{a}} ^{\; a}$ are assumed to be tangential to
the worldvolume, Eq. (\ref{Eua=ea}), and linearly independent, these
two are orthogonal to it, so that, taking into account the
superembedding equation (\ref{SembEq}), $E^i:=
E^{\underline{a}}u_{\underline{a}} ^{\; i}=0$. As it was first
noticed in \cite{bpstv}, this gives an equivalent representation of
the superembedding equation (\ref{SembEq}). It is convenient to
collect $u_{\underline{a}} ^{\; i}=(u_{\underline{a}} ^{\; 8}\, , \,
u_{\underline{a}} ^{\; 9})$ in two complex vectors
\begin{eqnarray}\label{uz=}
u_{\underline{a}} ^{\; z} = u_{\underline{a}} ^{\; 8}+ i
u_{\underline{a}} ^{\; 9} =( \bar{u}_{\underline{a}} ^{\;
\bar{z}})^*  \; , \qquad u_{\underline{a}} ^{\; z}u^{\underline{a}\;
b}=0 \; , \qquad u_{\underline{a}} ^{\; z}u^{\underline{a}\; z}=0 \;
,  \qquad \bar{u}_{\underline{a}} ^{\; \bar{z}}u^{\underline{a}\;
z}=-2 \; , \quad
\end{eqnarray}
so that the above equivalent form of the superembedding equation
reads
\begin{eqnarray}
\label{Ez=0} \hat{E}^{z}:= \hat{E}^{\underline{a}}
u_{\underline{a}}{}^z =0\; , \qquad \hat{E}^{\bar{z}}:=
\hat{E}^{\underline{a}} \bar{u}_{\underline{a}}{}^{\; \bar{z}} =0
  \; .
\end{eqnarray}

Actually, the superembedding equation (\ref{SembEq}) and the
conventional constraint (\ref{Eua=ea}) can be collected together in
the expressions for the pull--back of the type IIB bosonic
supervielbein form,
\begin{eqnarray}\label{Eua=eau} &&  \hat{E}^{\underline{a}} := {E}^{\underline{a}}(\hat{Z}(\zeta)) =
e^bu_b{}^{\underline{a}}(\zeta) \;  , \qquad
u^{\underline{a}b}u_{\underline{a}} ^{\; c}=\eta^{bc}\; , \qquad
\cases{ {\underline{a}} \; =\; 0,1,\ldots , 9\; , \cr b,c=0,1,\ldots
, 7\;  .}
   \qquad
\end{eqnarray}
This equation can also be obtained by substitution of the original
form of the superembedding equation, Eq. (\ref{SembEq}), into the
general decomposition of Eq. (\ref{hEa=b+f}) with $A=\underline{a}$.
Then the only information which is explicit in (\ref{Eua=eau}), in
comparison to the previously described equation, is the
orthogonality and normalization of the coefficient matrices,
$\hat{E}_a^{\,\underline{\,a}}= u_a^{\,\underline{\,a}}$. This
corresponds to the conventional constraints of choosing the bosonic
worldvolume supervielbein to be induced by (super)embedding, Eq.
(\ref{Eua=ea}).

\subsection{Fermionic supervielbein induced by superembedding and spinor moving frame}

To specify the induced supervielbein (\ref{eA=ea+}) of the
worldvolume superspace, we need, besides (\ref{Eua=ea}), to express
the $16$ fermionic forms $e^q$ (carrying the pseudo-Majorana spinor
representation of $SO(1,7)$, $(e^q)^*=\gamma^0_{qp}e^p$) in terms of
pull--backs of $32$ fermionic forms (\ref{Eua-cE}) of the target
type IIB superspace, ${\cal E}^{\underline{\alpha}}={\cal E}^{\alpha
i}=(E^{\alpha 1}\, , \, E^{\alpha 2})$. These latter carry the
spinor indices of the Majorana representation of $Spin(1,9)$.

To write such a fermionic conventional constraints we need in a
'bridge' (\cite{GIKOS})  between $Spin(1,7)$ and $Spin (1,9)$
groups, which is to say, in a matrix variable carrying one
$Spin(1,7)$ and one $Spin (1,9)$ indices. Such a bridge is given by
{\it spinor moving frame} matrix
\begin{eqnarray}\label{VinS7}
 V_\alpha{}^q \; \in Spin(1,7)\; , \qquad (V_\alpha{}^q)^* = \gamma^0_{qp}V_\alpha{}^q\;
 \qquad
\end{eqnarray}
providing a square root of the moving frame variables  in the sense of that
\begin{eqnarray}\label{su=VsV}
 &&  V_\alpha \gamma^a V_\beta
:=V_\alpha{}^q \gamma^a_{qp} V_\beta{}^p =
\sigma^{\underline{b}}_{\alpha\beta} u_{\underline{b}}{}^{\!{a}} \;
, \qquad \nonumber
\\  && V_\alpha{}^q (\delta+ i\gamma^9)_{qp}  V_\beta{}^p =
\sigma^{\underline{b}}_{\alpha\beta} u_{\underline{b}}{}^{\!{z}}\; ,
\qquad \qquad \nonumber
\\  &&
V_\alpha{}^q (\delta- i\gamma^9)_{qp}  V_\beta{}^p=
\sigma^{\underline{b}}_{\alpha\beta}
\bar{u}_{\underline{b}}{}^{\!\bar{z}}\; , \qquad
\end{eqnarray}
  as well as
\begin{eqnarray}
 \label{VTsV=}
V_q{\sigma}^{\underline{a}}V_p = \gamma^b_{qp}
u_{b}{}^{\!\underline{a}} - {1\over 2} (\delta+ i\gamma^9)_{qp}
\bar{u}^{\underline{a}\bar{z}} - {1\over 2} (\delta- i\gamma^9)_{qp}
u^{\underline{a} \, {z}}\; \; ,
\end{eqnarray}
where $\gamma^a_{qp}$ are $d=8$ gamma matrices and
$\sigma^{\underline{b}}_{\alpha\beta}$ are  the real $16\times 16$
sigma-matrices of the $SO(1,9)$ (see Appendix A; more details on
the moving frame variables can be found in Appendix C).

Using this spinorial moving frame matrix one can convert the
pull--backs of the fermionic target space supervielbein forms into
the one--forms with the $SO(1,7)$ spinorial index,
\begin{eqnarray}\label{Eq1=}
\hat{E}^{q 1}:= \hat{E}^{\alpha 1} V_{\alpha}{}^q\; , \qquad
(\hat{E}^{q 1})^*= \gamma^0_{qp} \hat{E}^{p 1}\; , \qquad \\
\label{Eq2=}
 \hat{E}^{q 2}:= \hat{E}^{\alpha 2} V_{\alpha}{}^q\; , \qquad
(\hat{E}^{q 2})^*= \gamma^0_{qp} \hat{E}^{p 2}\; . \qquad
\end{eqnarray}
The worldsheet fermionic supervielbein form $e^q$ can, in principle,
be identified with any of these two one-forms, or with a linear
combination of them. For D7-brane a convenient  conventional
constraint has  the form
\begin{eqnarray}\label{Eq1=eq}
\hat{E}^{q 1}:= e^q  \;  . \qquad
\end{eqnarray}
Then the general form of the second fermionic supervielbein is
\footnote{\label{footnote4} To describe an SD$7$-brane, which is
related to the standard D$7$-brane by a certain $SL(2)$ transformation, it is
convenient, following \cite{DpSL2}, to use  the counterpart of
(\ref{Eq1=eq}) imposed on the $SL(2)$ transformed doublet $(E^1,
E^2)$. Namely, in the notation of \cite{DpSL2},
\begin{eqnarray}
\label{Eqi=evi+} \nonumber  \hat{E}^{q i}:= e^q v^i + (e^ph_p{}^q +
 e^a \chi_a{}^q )  \tilde{v}^i \; , \qquad \tilde{v}^i =\epsilon^{ij} v_j \; , \qquad
 i,j=1,2\; , \qquad
\end{eqnarray}
where ${v}^i$ is the vector constructed from the axion and dilaton,
and $\tilde{v}^i =\epsilon^{ij} v_j$ is its complementary vector. We
however, will not use this constraint in the present paper; for our
purposes here the mere fact of the existence of the SL(2) covariant
formalism of \cite{DpSL2} will be sufficient.}
\begin{eqnarray}
\label{Eq2=eh+}
 \hat{E}^{q 2}:= e^p h_p{}^q + e^a \chi_a{}^q \; = \;   \hat{E}^{p 1} h_p{}^q + \hat{E}^a
 \chi_a{}^q\;  \; . \qquad
\end{eqnarray}

\subsection{Consequences of the superembedding equation}

The consistency conditions for the superembedding equation
(\ref{Ez=0}) read $d\hat{E}^z=0$. To write them in the covariant
form, it is convenient to introduce, besides the superspace spin
connection $w^{\underline{a}\underline{b}}:= dZ^{\underline{M}}
w_{\underline{M}}^{\underline{a}\underline{b}}(Z)$,  also the
worldvolume connections for the $SO(1,7)$ and $SO(2)=U(1)$ gauge
symmetries, $\omega^{ab}:=d\zeta^{{\cal M}}\omega_{{\cal
M}}{}^{ab}(\zeta)$ and $A:= \zeta^{{\cal M}}A_{{\cal M}}(\zeta)$.
These can be constructed  with the use of the spinor moving frame
variables in such a way that the covariant derivatives of the
orthogonal and tangential moving frame vectors read (see
\cite{bpstv,B00,Dima99} for more discussion in other examples)
\begin{eqnarray}
\label{Du=} D{u}_{\underline{a}} ^{\; {z}}= {u}_{\underline{a}b}
\Omega^{b {z}}\; , \qquad D\bar{u}_{\underline{a}}^{\; \bar{z}}=
{u}_{\underline{a}b} \bar{\Omega}{}^{b \bar{z}}\; , \qquad
D{u}_{\underline{a}} ^{\; b} ={1\over 2}{u}_{\underline{a}} ^{\;
{z}}\bar{\Omega}{}^{b \bar{z}} + {1\over 2}
\bar{u}_{\underline{a}}^{\; \bar{z}}\Omega^{b {z}} \; . \qquad
\end{eqnarray}
The 1-forms $\Omega^{b {z}}, \bar{\Omega}{}^{b \bar{z}}$ generalize (to the case
of curved target superspaces) the Cartan forms corresponding to the
$SO(1,9)/[SO(1,7)\times SO(2)]$ coset.

\subsubsection{Generalized Cartan forms,  Peterson--Codazzi, Gauss and
Ricci equations}

Eqs. (\ref{Du=}) provide us with  conventional constraints,
$\bar{u}^{\underline{a} \bar{z}}D{u}_{\underline{a}} ^{\; {z}}=0$
and ${u}^{\underline{c} a}D{u}_{\underline{a}}^{\; b}=0$, expressing
the worldvolume $U(1)=SO(2)$ and $SO(1,7)$ connections, $A$ and
$\omega^{ab}$, in terms of the pull--back
$\hat{w}^{\underline{a}\underline{b}}:=
w^{\underline{a}\underline{b}}(\hat{Z})$  of the target space spin
connection $w^{\underline{a}\underline{b}}(Z):= dZ^{\underline{M}}
w_{\underline{M}}^{\underline{a}\underline{b}}(Z)$ and moving frame
variables $u$ (entering through the true Cartan
forms $u^Tdu$).

Similarly to (\ref{Du=}), the covariant derivative of the spinor
moving frame matrix, which is the element
of $Spin(1,9)$ covering the $SO(1,9)$ matrix $U$, Eq.
(\ref{su=VsV}), is given in terms of the same generalizations of the
Cartan forms by
\begin{eqnarray}
 \label{DV=}
 DV_{\alpha}{}^{q} &=& {1\over 4} \; \Omega^{az}\;
V_{\alpha}{}^{p}(\gamma_a(\delta + i\gamma^9))_{pq} + {1\over 4} \;
\bar{\Omega}^{a\bar{z}} \; V_{\alpha}{}^{p}(\gamma_a(\delta -
i\gamma^9))_{pq}  \; . \qquad
\end{eqnarray}
This relation, which expresses the local  isomorphism of the SO(1,9)
and Spin(1,9) groups, can be derived by solving the equation
obtained by covariant differentiation of (\ref{su=VsV}) with the use
of (\ref{Du=}).

The selfconsistency conditions for Eqs. (\ref{Du=}) give the
following curved space generalization of the Maurer--Cartan
equations
\begin{eqnarray}
\label{MC=R} D\Omega^{a {z}} = (u\hat{R}u)^{a {z}}:=
R^{\underline{c}\underline{b}}{u}_{\underline{c}} ^{\;
{a}}{u}_{\underline{b}} ^{\; {z}} \; , \qquad D\bar{\Omega}{}^{a
\bar{z}}= (uRu)^{a \bar{z}}\; , \qquad
\end{eqnarray}
as well as   the following expressions for the $U(1)$ and $SO(1,7)$
curvatures
\begin{eqnarray}
\label{R=MC} r^{ab}:= d\omega^{ab} - \omega^{ac}\wedge \omega_c{}^b
= (uRu)^{ab} + \Omega^{[a| {z}}\wedge \bar{\Omega}{}^{|b]\bar{z}}\;
, \; \nonumber \\
dA= {i\over 2} (uRu)^{\bar{z}z} + {1\over 2} \Omega^{a {z}}\wedge
\bar{\Omega}_a{}^{\bar{z}}\; , \qquad
\end{eqnarray}
Eqs. (\ref{MC=R}), (\ref{R=MC}) involve the pull--back
$\hat{R}^{\underline{a}\underline{b}}:=
R^{\underline{a}\underline{b}}(\hat{Z})$ of the curvature
$R^{\underline{a}\underline{b}}:=(dw-w\wedge
w)^{\underline{a}\underline{b}}$ of the targets superspace spin
connections $w^{\underline{a}\underline{b}}$. They are the
supersymmetric --and also curved superspace-- generalizations of the
Peterson--Codazzi, Ricci and Gauss equations \cite{bpstv} written
for the case of codimension 2 supermebedding.

\subsubsection{Consequences of the superembedding equation}

Now we are ready to study the consistency conditions for the
superembedding equation (\ref{Ez=0}). Their covariant form is given by
$D\hat{E}^z=0$ and the complex conjugate equation
$D\hat{E}^{\bar{z}}=0$. Using Eq. (\ref{Du=}) and the bosonic
torsion constraints (\ref{Tua=}),
\begin{eqnarray}
\label{Tub=} {T}^{\underline{a}} := D{E}^{\underline{a}}= -i
(E^{{\alpha}1} \wedge E^{{\beta}1} - E^{{\alpha}2} \wedge
E^{{\beta}2}) \sigma^{\underline{a}}_{{\alpha}{\beta}}=: -i
E^1\wedge \sigma^{\underline{a}} E^1 - i E^2\wedge
\sigma^{\underline{a}} E^2 \; ,
\end{eqnarray}
one finds that the coefficient $h_p{}^q$ in the decomposition of the
pull-back of the second fermionic supervielbein form $E^{\alpha 2}$
obeys
\begin{eqnarray}
\label{hg9h=} h (\delta \pm i\gamma^9)h^T = - (\delta \pm i\gamma^9) \;
 \qquad
\end{eqnarray}
and also that the curved space generalizations of the ${SO(1,9)\over
SO(1,7)\times SO(2)}$ Cartan forms (\ref{Du=}) have the form of
\begin{eqnarray}
\label{Omz=} \Omega^{az} = - 2i e^q [h(\delta + i\gamma^9)\chi^a]_q
+ e_{b} (K^{ba\; z} - i \chi^b(\delta + i \gamma^9)\chi^a) \; ,
\qquad K^{ab\; z}=  K^{(ab)\; z}\;
\end{eqnarray}
and of its complex conjugate. Notice that, substituting (\ref{Omz=})
into the Peterson-Codazzi equation (\ref{MC=R}), one finds that its
lower dimensional component reads
\begin{eqnarray}
\label{DOmz=2} D_{(p}(h(\delta + i\gamma^9)\chi_a)_{q)} + {1\over
2}(\gamma^{b}+ h\gamma^bh^T )_{pq}(K_{ba}{}^{z}- i\chi_b(\delta +
i\gamma^9)\chi_a)= {i\over 4 }(\hat{R}^{az})_{pq} \; ,
\end{eqnarray}
where $(\hat{R}^{az})_{pq}=
(\hat{R}^{\underline{a}\underline{b}})_{pq}
u_{\underline{a}}{}^au_{\underline{b}}{}^z $ and
$\hat{R}^{\underline{a}\underline{b}}_{pq}$ appears as a lowest
dimensional coefficient in  the decomposition of the pull--back of
Riemann curvature two form on the basis of worldvolume
supervielbein, $\hat{R}^{\underline{a}\underline{b}}:= {1\over
2}\hat{E}^{\underline{D}}\wedge \hat{E}^{\underline{C}}
{R}_{\underline{C}\underline{D}}{}^{\!\!\underline{a}\underline{b}}(\hat{Z})
= {1\over 2}e^q \wedge e^p
(\hat{R}^{\underline{a}\underline{b}})_{pq}+ \propto e^a$. It is
expressed through pull--backs of the tangent superspace fluxes to
the worldvolume superspace. We will not need the explicit form of
$(\hat{R}^{az})_{pq}$ in this paper.

Eq. (\ref{DOmz=2}) relates the derivative of the fermionic
superfield $\chi_{a}{}^q= \hat{E}_a^{\underline{\alpha} 2}
V_{\underline{\alpha}}{}^q$ with the bosonic symmetric tensor
superfield
\begin{eqnarray}
\label{Kab:=}  K_{ab}{}^z=K_{ba}{}^z:=
-D_{(a}\hat{E}_{b)}{}^{\underline{a}}\, u_{\underline{a}}{}^z\; ,
\end{eqnarray}
which enters Eq. (\ref{Omz=}). Writing this in terms of
supervielbein pull--back, we have used the conventional constraint
(\ref{Eua=ea}) in its equivalent form  of
$u_a{}^{\underline{b}}=E_a{}^{\underline{b}}$.

In the purely bosonic case, when fermions are equal to zero, $K_{ab}{}^z$ is
called the {\it second fundamental form} of the bosonic worldvolume
$W^{8}$ embedded into the $D=10$ spacetime. Its trace  $h^z:= K_{a}{}^{az}$ is called {\it mean
curvature}. The vanishing of the main curvature of a bosonic surface
embedded in a flat space of higher dimension implies that this
surface is {\it minimal}. This also expresses, in terms of {\it
extrinsic geometry}, the equation of motion which follow form the
Nambu--Goto action. (Which explains the terminology: minimal surface
is obtained by minimizing the area (volume) of the surface
(hypersurface)).

In the case of flat superspace and vanishing worldvolume gauge
fields, the equations of motion of a superbrane also imply the
vanishing of the {\it mean curvature}, $h^z:= K_{a}{}^{az}=0$.
Clearly, in the case of generic curved superspace and of branes with
additional worldvolume gauge fields, bosonic equations of motion for
the scalar Goldstone (super)fields of the branes should be given by
a nonlinear generalization of this, including the pull--backs of
target superspace fluxes and also the worldvolume gauge field
contributions. Furthermore, Eq. (\ref{DOmz=2}) shows that such a
bosonic equation can be obtained as higher component of the
superfield fermionic equation, which in the week field limit (or for
the simplest case of membrane in flat tangent superspace) has the
form of Dirac equation
$\gamma^a_{qp}\chi_a{}^p=0\;$\footnote{Remember that
$\chi_a{}^p=\nabla_a \hat{Z}^M E_M^{\;\underline{\,\alpha}\,2}
V_{\underline{\alpha}}{}^p= \nabla_a \theta^{p\, 2}+ \ldots =
\nabla_a W^q\gamma^9_{qp}+ \ldots$; see sec. \ref{3.2.1}.
Furthermore, ignoring the products of fields and the flux
contributions, one finds that the linearized and flat superspace
version of Eq. (\ref{DOmz=2}) implies $ K_a{}^{az}=- {i\over
16}D_q(\gamma^a\chi_a)_q$. }. So, as is usual in supersymmetric
theories, it is sufficient to find the superfield fermionic equation
and then the bosonic scalar equations will appear in its
decomposition in the Grassmann coordinates.

However, neither such a fermionic  equation nor bosonic equations of
motion appear as a consequences  of superembedding equation
(\ref{Ez=0}) for the IIB super-7--brane.

Thus to describe the type IIB 7--brane dynamics we have to search
for additional constraints which would lead to the equations of
motion. \footnote{In the light of the off-shell nature of the
superembedding equation for type IIB 7-branes, one could also search for a superfield action
of STV type (see \cite{stv,DGHS93} and \cite{Dima99} for review and
further references) producing these equations of motion together
with the superembedding equations. We do not try to elaborate this
direction in the present paper.}

\subsubsection{Selfconsistency of the equations for fermionic supervielbein forms}

%%%%%270109
One should also study selfconsistency conditions for the fermionic equations
(\ref{Eq2=eh+}), (\ref{Eq1=eq}). Although  (\ref{Eq1=eq}) is a conventional constraint  and (\ref{Eq2=eh+}) just manifest the general decomposition of the second fermionic supervielbein form on the basis of the worldvolume superspace one forms, such integrability conditions provide us with the properties of $h_p{}^q$ and $\chi_a^q$ superfields which they possesses by their definition in Eqs. (\ref{Eq2=eh+}), (\ref{Eq1=eq}). To clarify this point, let us notice that in the linearized approximation and in flat superspace $h_{p}{}^q=D^0{}_{\!\! p}\, \theta^{2 q}$ and $\chi_{a}{}^q=\partial_a \theta^{2 q}$ and, hence,  $D{}^0{}_{\!\! (p}\, h_{p^\prime )}{}^q
=2i \gamma^a_{pp^\prime } \chi_{a}^q$ and $\partial_{[a}\chi_{b]}{}^q=0$. The selfconsistency  conditions for (\ref{Eq2=eh+}), (\ref{Eq1=eq}) carry the nonlinear counterpart of these two equations valid in an arbitrary superspace supergravity background.

In particular, the lowest, 1/2 dimensional component of the integrability condition $D(E^{2q}- e^ph_p{}^q -e^a\chi_a^q)=0$ for Eq. (\ref{Eq2=eh+}) produces the expression for $D_{(p}h_{p^\prime)}{}^{q}$ (see Eq. (\ref{Dphpq=}) in Appendix D), the trace part of which reads
\begin{eqnarray}
\label{trDphpq=} D_{p}h_{p}{}^{q} &=  -14\left( (hV)_{q}{}^\alpha
\widehat{D_{\alpha 1}e^{-\Phi}} + V_{q}{}^\alpha
\widehat{D_{\alpha 2}e^{-\Phi}}\right) \; ,  \qquad
\end{eqnarray}
where $\widehat{D_{\alpha 1,2}e^{-\Phi}}$ are the pull--backs   of $D_{\alpha 1,2}e^{-\Phi}$ superfield to $W^{(8|16)}$. Below we will omit hat symbols from the pull--backs in the places where this cannot produce a confusion.

Eq. (\ref{trDphpq=}) is not dynamical when considered together with superembedding equation only. However, as we will discuss below, after imposing the D7-brane gauge field constraint an additional algebraic equation for $h$ appears (see Eq. (\ref{hgh=gk}) below). Considered together with this,  Eq. (\ref{trDphpq=}) becomes dynamical and, moreover, collecting all the set of the dynamical equations of motion.

\subsection{Superspace constraints for D7-brane worldvolume gauge field }

In the case of M-branes, D=10 fundamental string and D$p$--branes
with $p\leq 5$ the superembedding equation contains equations of
motion among their consequences \cite{bpstv,hs2,HSW+Chu=PLB98}.
Hence, the description of {\it the set of all possible p--branes} by
this equation is complete. Then what happens if several type of
$p$-branes are possible? In \cite{B00} this question was addressed
for the strings (1-branes) in type IIB superspace; it was shown
there that the superembedding equation provides a universal
description of fundamental string (sometimes called F1--brane) and
Dirichlet 1-brane (D1-brane) \footnote{The generic solution of the
superembedding equation for $p=1$ describes the case of
super--D1--brane, and the worldvolume gauge filed strength enter the
solution as a parameter. In the simplest case of flat superspace
this is just a constant parameter, the appearance of which had been
observed already in \cite{bpstv}. From the side of this generic
D1-brane case, the solution describing the fundamental string
appears in the limit when the D1-brane field strength reaches its
maximal possible value \cite{B00}. Notice that another universal
description of string and D$1$--brane is provided by SL(2) covariant
superstring action of \cite{MC+PKT1997}. }. Similarly, the
superembedding equation can provide the universal description of all
possible type IIB 5-branes, so that the search for possible
existence of new 5-branes (if any) can be performed by searching for
possible existence of new solutions of the superembedding equation,
different from the ones describing D5-brane and its SL(2) images
(including NS5; these are called $(p,q)$ five branes, although we
could also propose the name of SD5-brane).

This is not the case when $p=7$. The superembedding equation for
this case is not dynamical and, to describe the dynamics of
D7--brane, the constraints on the worldvolume gauge field strength
(\ref{F2=dA-B2})  should be imposed \cite{HSW+Chu=PLB98}. This
constraint  reads
\begin{eqnarray}\label{F2=dA-B}
{\cal F}_2 = dA-\hat{B}_2 = {1\over 2} e^b\wedge e^a F_{ab}\; ,
\qquad
\end{eqnarray}
where $e^a$ is the worldvolume superspace bosonic vielbein induced
by superembedding, Eq. (\ref{Eua=ea}), and $\hat{B}_2$ is a
counterpart of the WZ term of the fundamental string, but on the
7-brane worldvolume superspace. In general curved type IIB background
this is given by the pull--back of the NS--NS two--form gauge field
$B_2$, the field strength of which, $H_3=dB_2$, is restricted by the
following superspace constraints
\begin{eqnarray}
\label{H3=} && H_{3}:=dB_2 = - i {E}^{\underline{a}}\wedge
({E}^{1}\wedge \sigma_{\underline{a}} {E}^{1} -  {E}^{2}\wedge
\sigma_{\underline{a}} {E}^{2}) +{1\over 3!}
{E}^{\underline{c}_3}\wedge{E}^{\underline{c}_2}\wedge
{E}^{\underline{c}_1}
H_{\underline{c}_1\underline{c}_2\underline{c}_3}\; . \qquad
\end{eqnarray}
In the case of flat type IIB superspace, the last, bosonic 3-form
contribution vanishes, and Eq. (\ref{H3=}) is equivalent to the
first component of the doublet equation (\ref{F3R=flat}), as it is
indicated by (\ref{dF2R=(H,R)}).

Now, studying the Bianchi identities (\ref{F2=dA-B2}) for the
worldvolume field strength one finds, besides the purely bosonic
identity $3D_{[a}F_{bc]} =-H_{abc} - 6i (\eta + F)_{d[a}\,
\chi_{b}\gamma^d\chi_{c]}$ and
\begin{eqnarray}
 \label{DqF=}
 D_qF_{bc}= - 4i (\eta -  F)_{a[b}
(h\gamma^a\chi_{c]})_q\; ,
\end{eqnarray}
that the spin-tensor $h_p{}^q$ in (\ref{Eq2=eh+}) obeys the algebraic equation
\begin{eqnarray}
\label{h(1-F)h=(1+F)} h\gamma^ah^T (\eta -F)_{ab} = \gamma^a (\eta
+F)_{ab}  \; . \qquad
\end{eqnarray}
For $F_{ab}=0$ the system of this equation and Eq. (\ref{hg9h=})  is
solved by $h\!\!^{^0}{}_p{}^q = (\gamma^9)_{pq} $. In the generic
case of nonvanishing $F_{ab}$, Eq. (\ref{h(1-F)h=(1+F)}) can be
written as
\begin{eqnarray}
\label{hgh=gk}  h\gamma^ah^T = \gamma^b k_b{}^a\;  , \qquad
k_b{}^a:=
 (\eta
+F)_{bc}(\eta- F)^{-1}{}^{ca} \; \in \; SO(1,7)\; . \qquad
\end{eqnarray}
This includes the pseudo--orthogonal $8\times 8$ matrix $k_a{}^b$
($k\eta k^T=\eta$) the Cayley image of the antisymmetric tensor
$F_{ab}$. Eq. (\ref{h(1-F)h=(1+F)}) implies that $h$ is an element
of the $Spin(1,7)$ group. Eqs. (\ref{h(1-F)h=(1+F)}) and
(\ref{hg9h=}) are solved by
\begin{eqnarray}\label{h8=}
  h_p{}^q  = & {1\over \sqrt{|\eta +F|}} \left[\gamma^9 + {1\over
  2}F_{ab}\gamma^{ab}\gamma^9 - {1\over 8\cdot 4!} (\varepsilon
  FF)^{a_1a_2a_3a_4}\gamma_{a_1a_2a_3a_4}  + \right. \qquad
  \nonumber \\ &\left. \qquad +
{1\over 4\cdot 4!} (\varepsilon
  FFF)^{ab}\gamma_{ab} - {1\over 16\cdot 4!} (\varepsilon
  FFFF)\; \delta\; \right]{}_{pq} \; . \qquad
\end{eqnarray}

More discussion on the Lorentz group valued spin tensor $h$ for different D$p$--branes can be found in \cite{ABKZ,IB+DS06}.
Special properties of the $p=7$ case are related to that, due to (\ref{hg9h=}),
\begin{eqnarray}\label{hhT=-1}
h^Th=-I=hh^T \, , \qquad h\gamma^9h^T=-\gamma^9\; ,
\;  \qquad
\end{eqnarray}
so that the matrix inverse to $h$ is given by $-h^T$. This implies
\begin{eqnarray}\label{hgnhT=}
h\gamma^{a_1\ldots a_n}h^T=(-)^{n+1} \gamma^{b_1\ldots b_n}k_{b_1}{}^{a_1}\ldots k_{b_1}{}^{a_n} \, , \quad h\gamma^{a_1\ldots a_n}\gamma^9h^T=(-)^{n} \gamma^{b_1\ldots b_n}\gamma^9\, k_{b_1}{}^{a_1}\ldots k_{b_1}{}^{a_n} \, . \quad
\end{eqnarray}

\subsection{D$7$--brane equations of motion from gauge field
constraints plus superembedding equations}

Eq. (\ref{hgh=gk}) implies that $h^{-1}dh$ takes values in $spin(1,7)$ Lie algebra,
$k^{-1}dk$ takes its values in $so(1,7)$ and
\begin{eqnarray}
 \label{h-1Dh=kTDk} h^{-1}Dh = {1\over 4}(k^{-1}Dk)^{ab}\gamma_{ab}\; .
 \end{eqnarray}
Then, as far as, by construction, $(k^{-1}Dk)^{ab}= 2(\eta + F)^{-1}DF\,
(\eta - F)^{-1}$, (\ref{h-1Dh=kTDk}) implies
\begin{eqnarray}
 \label{h-1Dh=} h^{-1}Dh &=& \;\; {1\over 2}\, DF_{ab}(\eta- F)^{-1}{}^{ac}(\eta- F)^{-1}{}^{bd}\gamma_{cd}\; , \qquad \\
\label{Dhh-1=}
Dh\, h^{-1} &=& - {1\over 2}DF_{ab}(\eta + F)^{-1}{}^{ac}(\eta + F)^{-1}{}^{bd}\gamma_{cd}
\; . \qquad
 \end{eqnarray}
Eq.  (\ref{Dhh-1=}) is obtained from (\ref{h-1Dh=}) using Eq. (\ref{hgnhT=}) with $n=2$.

Now, substituting (\ref{DqF=}) into the fermionic component of Eq. (\ref{Dhh-1=}),  one finds
$(D_qh\, h^{-1})_{ p^\prime q^\prime}= - 2i (h\gamma^a \chi_c)_q (\eta- F)^{-1}{}^{cb}\, (h\gamma_{ab}h^T)_{p^\prime q^\prime}$. Taking into account that $h^T=-h^{-1}$, one can present this equation in the form of
%(remember that $h^T=-h^{-1}$),
 $(D_qh)_{ p^\prime p}= 2i (h\gamma^a \chi_c)_q (\eta- F)^{-1}{}^{cb}\, (h\gamma_{ab})_{ p^\prime p}$; the trace of this gives
\begin{eqnarray}
 \label{Dh=gchi} D_{q}h_{q}{}^{p} &=& -14i  (\eta- F)^{-1}{}^{ab}(\gamma_{b}\chi_a)_p
\; . \qquad
 \end{eqnarray}
However, $D_{q}h_{q}{}^{p}$ is expressed in terms of pull--backs of background superfields by the consequence (\ref{trDphpq=}) of Eqs. (\ref{Eq2=eh+}), (\ref{Eq1=eq}). Thus we arrive at
\begin{eqnarray}
 \label{fEqD7=}
 (\eta -F)^{-1}{}^{ab}\gamma_{b}{}_{qp}\chi_a{}^p =-i \left( (hV)_{q}{}^\alpha
{D_{\alpha 1}e^{-\Phi}} + V_{q}{}^\alpha
{D_{\alpha 2}e^{-\Phi}}\right)  \; ,
\qquad \chi_a^p:= \hat{E}_a^{\alpha 2}V_\alpha{}^p\;\;
 \qquad
 \end{eqnarray}
which is the (superfield generalization of the) fermionic equation of motion for
D$7$--brane\footnote{See \cite{bst97,ABKZ,IB+DS06} for the
D$p$--brane equations and earlier \cite{hs2} for the M5--brane
case.}.

In the {\it flat D=10 type IIB superspace} $D_{\alpha 1}e^{-\Phi}=0=D_{\alpha 2}e^{-\Phi}$, so that the {\it r.h.s}'s of both Eq. (\ref{fEqD7=}) and Eq. (\ref{trDphpq=}) vanish. Hence the latter equation simplifies  to $D_qh_q{}^p=0$ and  Eq. (\ref{fEqD7=})
can be written as
\begin{eqnarray}
 \label{fEqD7=flat} \gamma_{a}{}_{qp} (\eta +
F)^{-1}{}^{ab}D_b\Theta^{2\,p}=0\;
 \qquad
 \end{eqnarray}
(notice that $(\eta- F)^{-1}{}^{ba}= (\eta
+F)^{-1}{}^{ab}$) . This differs from the standard Dirac equation by the contribution of the
worldvolume flux $F_{ab}$: decomposing (\ref{fEqD7=flat}) in power series on $F$ one easily finds $\gamma^{a}{}_{qp} D_a\Theta^{2\,p}= \gamma_{a}{}_{qp}
F^{ab}D_b\Theta^{2\,p} + {\cal O}(FF)$.

As usually, the bosonic equations of motion can be
obtained by acting on the fermionic equations by covariant spinor
derivative. On this way, one should use also the algebraic
consequences of the superembedding equations, including Eq.
(\ref{DOmz=2}). We will not need in the explicit form of the
D7--brane bosonic equations in our discussion below. See
\cite{IB+DS06} for detailed study of the (component form of
the) type IIB D$p$--brane equations.

\section{Searching for a superembedding description of Q7-brane}

The discussion in our Sec. \ref{Ch-E=Co} supports the conclusion of
\cite{Dima+Eric07} that the Q7--branes (if exist as dynamical
objects) should carry two gauge fields on their worldvolume.

As far as a complete supersymmetric description is concerned, the
problem with counting bosonic and fermionic degrees immediately
arise. In the case of D7-brane (and SD7-branes of \cite{DpSL2}) the
number of dynamical bosonic and fermionic degrees of freedom
coincide ($2+6 =16/2$) and the gauge field degrees  of freedom enter
the balance (as '$6$'). Now, adding a new gauge field, one should
either add for him additional fermionic degrees of freedom or to
assume its nondynamical/dependent nature. As a mechanism for the
latter, the authors of \cite{Dima+Eric07} suggested a possibility of
some generalization of duality equations. However, no mechanism to
generate such generalized duality equations is know. Let us note in
this respect that it is natural to assume that, if existed, such
description with self-duality equations should be also reproducible
in the frame of superembedding approach, like, for instance, the
self-duality of the five--form field strength follow from the
superspace constraints of type IIB supergravity \cite{HoweWest83} or
like the nonlinear generalization of the six dimensional self-duality equation
follows from the superembedding equation of M5-brane \cite{hs2} (see
\cite{hsw97,bpstvPLB} for details).

The above statement actually is based on more general conjecture:
{\it if existed, the Q7-brane should also allow for a superembedding
description}. The reason for this is that it was the case for all
previously known branes; more generally, to our best knowledge,
there is no one known example of a supersymmetric system (neither
field theoretical nor brane-type) which do not allow for an {\it
on-shell} superfield description.

Furthermore, it is also natural to assume that the Q7--brane
worldvolume superspace obeys the superembedding equation, Eq.
(\ref{SembEq}) or equivalently (\ref{Ez=0}). Again, the reason is
that  this is the case for the (complete) superfield description of
all the presently known 1/2 BPS superbranes (which means
super-$p$--branes the ground state of which preserves one half of
the target space supersymmetry), and the supergravity solutions
describing Q7--branes are 1/2 BPS
\cite{Eric+06sD7,Eric+06aD7}\footnote{See \cite{kappaInt} for the
relation between projectors of the $\kappa$--symmetry of
brane actions and supersymmetry preserved by corresponding
bosonic solutions of supergravity equations, \cite{BdAI1}  for
the supersymmetry preserved by bosonic brane actions (bosonic limit
of superbrane action) and \cite{IB+JdA05} for the complete but gauge
fixing Lagrangian description of the supergravity--superbrane
interacting system, which explains the above mentioned relation
between the supersymmetry and the $\kappa$--symmetry.}.

Then, the worldvolume superspace of the Q7--brane, if existed, would
carry two 'linearly independent' worldvolume gauge potential
super-1-forms, {\it i.e.} two super-1-forms with generalized field
strengths given by $q^R_1{\cal F}_R$ and $q^R_2{\cal F}_R$ with
linearly independent doublets of charges, $q^R_1$ and $q^R_2$, and
${\cal F}_R$ defined in (\ref{F2R=(H,R)}). Notice that, as we
discussed in Sec. \ref{2.3},  such a straightforward lifting to
worldvolume superspace is what is necessary for having a possibility
to construct the standard-type Q7-brane action by superembedding
approach based method of \cite{Howe-GAP}.

To simplify our study, let us  consider a particular situation when
one of two worldvolume gauge fields of  the hypothetical Q7-brane is
the familiar gauge field of D-brane, $A=d\zeta^M A_M(\zeta)$, related to the pull--back of
NS-NS form, (\ref{F2=dA-B}), \begin{eqnarray}\label{F2=dA-B2-2}
{\cal F}_2=dA- \hat{B}_2  \; , \qquad
\end{eqnarray}
while the other, which we call ${A}^{(q , q^\prime)}= d\zeta^M A^{(q , q^\prime)}_M(\zeta)$, has its field strength defined by
\begin{eqnarray}\label{G2=dA-BC}
G_2 :=  d{A}^{(q , q^\prime)} - q \hat{B}_2 - q^\prime \hat{C}_2\;
\;  \qquad
\end{eqnarray}
with some constants $q$ and $q^\prime\not= 0$. The existence of the
SL(2) covariant formalism \cite{DpSL2} guaranties that the general
situation can be reproduced by certain SL(2) transformation of the
above choice (see also comment in footnote \ref{footnote4}).

The field strength $G_2$ defined by Eq. (\ref{G2=dA-BC}) is
invariant, besides the abelian worldvolume gauge transformations,
under the NS-NS and RR target superspace gauge transformations,
\begin{eqnarray}\label{dA=(dBdC}
\delta {B}_2 =d{\alpha}_1\; , \qquad \delta {C}_2 =
d{\alpha}^\prime_1\; ,  \qquad \delta A^{(q , q^\prime)}= q
\hat{\alpha}_1 + q^\prime \hat{\alpha}^\prime_1 + d\alpha_0\; .
\end{eqnarray}
Similarly, the invariance of the field strength ${\cal F}_2$ is
guaranteed by that $\delta A=\hat{\alpha}_1$; this, in contrast to
generic $A^{(q,q^\prime)}$ with $q^\prime\not= 0$, is inert under the RR gauge transformations.

Now the question arise: what are the constraints which should be
imposed on these two field strengths?

\subsection{Candidate constraints for two worldvolume gauge
potentials on the worldvolume superspace}

{\it A strong suggestion comes form the linearized analysis}.
Indeed, if there exists a worldvolume model of Q7-branes with two
dynamical gauge fields, this should allow, in particular, for zero
values of the fields, and, hence, for a weak field approximation.
Thus it is reasonable to search first for the linearized description
of the two independent and different gauge fields on a 7-brane
worldvolume.

Then one can check that the following constraints
\begin{eqnarray}\label{F2=dA-B2=2}
{\cal F}_2 &=& {1\over 2} e^b\wedge e^a F_{ab} \\ \label{dA-B-C=}
G_2 &=& {1\over 2} e^b\wedge e^a G_{ab} + e^b\wedge e^q \gamma_{b\,
qp} {\cal W}^p  + {1\over 2} e^p\wedge e^q ((\delta + i
\gamma^9)_{pq}\Upsilon + (\delta - i \gamma^9)_{pq}\bar{\Upsilon})
\; , \qquad
\end{eqnarray}
have a correct weak field limit. Namely, they, together with the
linearized superembedding equations result in the Dirac equation
for the fermionic fields, Klein--Gordon equations for scalars and
Maxwell equations for the vector fields of two d=8 vector
multiplets,\footnote{It is also worth noticing that the set of
constraints (\ref{F2=dA-B2=2}), (\ref{dA-B-C=}) have the $SL(2)$
covariant generalization which, in notation of \cite{DpSL2}, reads
\begin{eqnarray}\label{F2=dA-B2=2c}\nonumber
{\cal F}^{(2)}_r = {1\over 2} e^b\wedge e^a {\cal F}_{ab\; r} +
\tilde{V}_r \left( e^b\wedge e^q \gamma_{b\, qp} {\cal W}^p  +
{1\over 2} e^p\wedge e^q ((\delta + i \gamma^9)_{pq}\Upsilon +
(\delta - i \gamma^9)_{pq}\bar{\Upsilon}) \right) \; , \qquad
\end{eqnarray}
where $\tilde{V}_r$ is a 2--dimensional $SO(2)$ vector constructed
from axion and dilaton, namely $\tilde{V}_r=\epsilon^{rs}
{V}_s$ with $V_r= \sqrt{2}
v^i (\tau_r)_{ij} \tilde{v}^j$,  and the SO(2) spinors
${v}^i$ and $\tilde{v}^i=\epsilon^{ij}v_j$ constructed from the
axion and dilaton; these are the ones used in an SL(2) covariant
formulation of the fermionic conventional constraints (\ref{Eq1=eq}) and (\ref{Eq2=eh+}), $\hat{E}^{q
i}:= e^q v^i + (e^ph_p{}^q +
 e^a \chi_a{}^q )  \tilde{v}^i$ \cite{DpSL2}, discussed  in
 footnote  \ref{footnote4}.}
 \begin{eqnarray}\label{DEq-Q}
 \hbox{Superembedding multiplet} \qquad
& | & \qquad \hbox{Additional vector multiplet} \qquad \nonumber \\
\label{MEq-Q}
\partial_{[c}F_{ab]} =0 \; , \qquad   \partial^b F_{ab} =0\; , \qquad
& | & \qquad   \partial_{[c}G_{ab]} =0 \; . \qquad   \partial^b
G_{ab} =0\; , \qquad \\ \label{FEq-Q}
\partial\!\!\!/_{pq}{W}^q =0 \; , \qquad & | & \qquad   \partial\!\!\!/_{pq}{\cal W}^q =0 \; ,
\qquad \\ \label{SEq-Q} \Box \tilde{X}^{z}=0  \; , \qquad  \Box
\tilde{\bar{X}}{}^{\bar z}=0 \; , \qquad & | & \qquad   \Box
\Upsilon =0 \; , \qquad \Box \bar{\Upsilon} =0 \; . \qquad
\end{eqnarray}
Here we separated the field equations in two blocks coorresponding
to two different vector multiplets. The first of these two d=8
vector multiplets is formed by leading components of the Goldstone
superfields of Eq. (\ref{XTh=fields}) and of the antisymmetric
tensor superfield $F_{ab}$ in (\ref{F2=dA-B2-2}). We call this the
{\it superembedding multiplet} and its equations of motion are
actually the linearized equations of motion of the
super--D$7$--brane. The second $d=8$ linearized vector multiplet is
formed by leading components  of the $G_{ab}$, $\Upsilon=
(\bar{\Upsilon})^*$ and ${\cal W}^q$ superfields in (\ref{dA-B-C=}).
In the linearized approximation this constraints is considered on
the flat $d=8$ superspace because the contributions to the
worldvolume geometry from the superembedding multiplet are neglected
as being of second order in fields.

It is a place to stress that, in the 'rigid' Q7--brane picture it is
hard to find a place for the 2 bosonic and 16 fermionic fields
$\Upsilon$ and ${\cal W}^q$ (we use the same notation for the
superfields and their leading components when this cannot produce a
confusion). On the other hand, such fields look quite natural if we
think about Q7--brane as about bound state of two SD7--branes (one
of which is identified, for simplicity, with D7--brane). Then these
additional fields  $\Upsilon$ and ${\cal W}^q$ complete the
additional bosonic field $G_{ab}$ up to a vector supermultiplet
which is identified as the superembedding supermultiplet of the
second SD7--brane.

The problem of superpartners of the second gauge field on the
hypothetical Q7--brane, have been noticed in \cite{Dima+Eric07}. To
escape the treatment of Q7 as a system of two SD$7$--branes it was
proposed there that the two gauge field strengths are related
by a kind of nonlinear d=8 generalization of the selfduality
condition, like, schematically, $*G\propto F\wedge F\wedge F$ or
$G\wedge G=
*(F\wedge F)$\footnote{More precisely, the equations proposed in \cite{Dima+Eric07} were
${\cal G}^-\wedge {\cal G}^-=
*{\cal G}^-\wedge {\cal G}^- $ with ${\cal G}^{\pm}=G\pm {\cal F}$, and ${\cal G}= *({\cal G}\wedge
*{\cal G}\wedge {\cal G}) $, but the explicit forms is not essential for our present study.}, so that the second set of superpartners is not
needed.

No dynamical mechanism for generating such a nonlinear generalized
duality equation was proposed in \cite{Dima+Eric07}. But, if
existed, such equation should follow from superembedding approach,
{\it i.e.} appear as a requirement of selfconsistency of the system
including superembedding equation and the gauge field constraints
{\it without} additonal superfields $\Upsilon$ and ${\cal W}^q$. If
this were the case, then the addition of unwanted bosonic and
fermionic (super)fields $\Upsilon$ and ${\cal W}^q$ cannot spoil the
results: these fields would either be set to zero by the above
mentioned selfconsistency conditions, or allow for being set to zero
at the final result\footnote{One might hope that these would be
auxiliary fields providing the off--shell extension of the first,
superembedding supermultiplet, but such a possibility  has been
actually excluded by the linear approximation analysis which results
in dynamical equations (\ref{MEq-Q}), (\ref{FEq-Q}),
(\ref{SEq-Q}).}. The same analysis also shows that the linearized
$G_{ab}$ vanishes if $\Upsilon=0$ and ${\cal W}^q=0$; this excludes
the possibilities of generating equations of the type of $G\wedge G=
*(F\wedge F)$, but still  allows to conjecture the appearance of
{\it e.g.} $*G=\propto F\wedge F\wedge F$ equation, as far as its
linearized limit would be just $G_{ab}=0$. This conjecture on
possible appearance of such a duality conditions can now be checked.

To this end we shall study the selfconsistency conditions for the
superembedding equation (\ref{Ez=0}) and the constraints
(\ref{F2=dA-B2=2}), (\ref{dA-B-C=}), the form of which have been
motivated by the consistency of the linearized approximation.
Actually, the consistency of the superembedding equations
(\ref{Ez=0}) and the constraints (\ref{F2=dA-B2=2}) for the first
worldvolume gauge field strength has been already checked in Sec.
\ref{SuperEmb}. This consistency requires the D7--brane equations of
motion to hold.

Thus {\it the search for a Q7--brane description in
the frame of superembedding approach is reduced to checking of
whether it is possible to have the second different gauge field
super-1-form potential on the worldvolume superspace of the
D7--brane}. Furthermore, it actually reduces  to checking of whether
the constraints (\ref{dA-B-C=}) can be consistent on such a
worldvolume superspace.

{\it The result of such a check is negative}. Taking into account
the importance of this conclusion, we present below some technical
details.

\subsection{Solving the Bianchi identities for the second gauge field
on the worldvolume superspace of D$7$--brane }

Using the consequences $\gamma^a + h\gamma^ah^T= 2\gamma_b(\eta
-F)^{-1}{}^{ba}\;$, and $\gamma^a - h\gamma^ah^T= -2\gamma^b\;
F_{bc}(\eta -F)^{-1}{}^{ca}\;$ of Eq.  (\ref{hgh=gk}), one finds
that the lowest dimensional (dim $3/2$) contribution to the Bianchi
identities
\begin{eqnarray}\label{dG=H+R}
dG_2+ (q + q^\prime \hat{C}_0)\hat{H}_3 + q^\prime \hat{R}_3  =  0
\;  \qquad
\end{eqnarray}
for the constraints (\ref{dA-B-C=}), (\ref{G2=dA-BC}) reads
\begin{eqnarray}\label{QdG(3/2)=0}
2i \gamma_{b \, (q_1q_2} (\eta -F)^{-1}{}^{ba} (\gamma_a{\cal
W})_{q_3)} +
 {1\over 2} (\delta + i\gamma^9)_{(q_1q_2}D_{q_3)} \Upsilon +
 {1\over 2} (\delta - i\gamma^9)_{(q_1q_2}D_{q_3)} \bar{\Upsilon} =
 \qquad \nonumber \\ = -
  {1\over 2} T_{(q_1q_2}{}^p \left(
 (\delta + i\gamma^9)_{q_3)p} \Upsilon +
(\delta - i\gamma^9)_{q_3)p} \bar{\Upsilon}\right) \; . \qquad
\end{eqnarray}
The explicit form of the worldvolume fermionic torsion $T_{pq}{}^{q^\prime}$ can be found in Appendix D (see Eq. (\ref{Tfff(w-sh)=}) and also (\ref{fppqq:=})).  Decomposing Eq. (\ref{QdG(3/2)=0}) on the irreducible parts, one finds (see Appendix E for details)
\begin{eqnarray}
\label{DY=gW+} & D_q \Upsilon = - 2i (\delta -
i\gamma^9)_{qp}\left({\cal W}{}^p  -\Upsilon \Lambda^1_{p} \right)\;
, \qquad
\end{eqnarray}
with
\begin{eqnarray} \label{L1=def}  \Lambda^1_{q} := {i\over
2}V_q{}^\alpha\widehat{(D_{\alpha 1}e^{-\Phi})}\; ,  \qquad \Lambda^2_{q} := {i\over
2}V_q{}^\alpha\widehat{(D_{\alpha 2}e^{-\Phi})}\; ,  \qquad
\end{eqnarray}
and also
\begin{eqnarray}\label{cUbp=0}
(F(\eta -F)^{-1})_b{}^{c} ((\delta + i\gamma^9)\gamma_c{\cal W})_{p}
- 2 ((\delta + i\gamma^9) \gamma_b\Lambda_1)_p\, {\Upsilon} -
((\delta + i\gamma^9) h\chi_b)_p\, \bar{\Upsilon}  =0 \; . \qquad
\end{eqnarray}
In distinction to (\ref{DY=gW+}), Eq.  (\ref{cUbp=0}) is essentially
nonlinear: it does not have a nontrivial linear approximation in the
case of flat background superspace.

One also finds the equations complex conjugate to (\ref{DY=gW+}) and
(\ref{cUbp=0}), so that it is possible to find, in particular, the expression
for $(F(\eta -F)^{-1})_b{}^{c} (\gamma_c{\cal W})_{q}$. However, it
is more convenient to use first the the dynamical equation for the
superembedding fermion, which is just the D7--brane fermionic
equation (\ref{fEqD7=}),
\begin{eqnarray}\label{fEqD7=2L}
(\eta- F)^{-1}{}^{ba} \gamma_a\chi_b=-2(\Lambda^{2}+h\Lambda^{1})  \; ,  \qquad
\end{eqnarray}
% in its equivalent form $(\eta- F)^{-1}{}^{ba}
%\gamma_ah\chi_b=2h(\Lambda^{2}+h\Lambda^{1})$, and also Eq.
%(\ref{hgh=gk}),  one can
and to write Eq. (\ref{cUbp=0}) in the equivalent form
\begin{eqnarray}\label{gcUbp=0}
(F(\eta -F)^{-2}){}^{ab}(\delta -
i\gamma^9)\gamma_{a}\gamma_{b}{\cal W} = 2(\delta -
i\gamma^9)\left(\Upsilon \gamma_{a}(\delta
-F)^{-1}{}^{ab}\gamma_{b}\Lambda^1 + \bar{\Upsilon}
h(\Lambda^2+h\Lambda^1)\right)\, . \;
\end{eqnarray}
In the derivation of (\ref{gcUbp=0}) one uses Eq. (\ref{hgh=gk}) and
(\ref{hhT=-1}). Notice that {\it r.h.s.} of this equation is
proportional to the pull--backs of spinorial derivatives of the
dilaton superfield  (\ref{L1=def}), which can be called {\it fermionic
fluxes},  and, hence, vanishes for the
7-brane in flat target superspace where (\ref{gcUbp=0}) and its complex
conjugate imply $(F(\eta -F)^{-2}){}^{ab} (\gamma_{a}\gamma_{b}{\cal
W})_{q} =0$. This equation stating vanishing of the product of
fields from different supermultiplets already suggests a possibility
of that these cannot coexist. But to see that this is indeed the case
we have to go through further studying the consequences of Eq.
(\ref{gcUbp=0}) and the dim $2$ component of the Bianchi identities
(\ref{dG=H+R}).

Acting by the fermionic derivative $D_p$ on (\ref{gcUbp=0}), one finds
\footnote{In derivation of Eq. (\ref{gcUbp=0+0}) one have also take into account Eqs. (\ref{DqF=}) and (\ref{h-1Dh=}).}
\begin{eqnarray}\label{gcUbp=0+0}
(F(\delta -F)^{-2}){}^{ab} (\gamma_{a}\gamma_{b})_{pq^\prime}
D_q{\cal W}^{q^\prime} + {\cal O}({\Upsilon})+ {\cal O}({\cal W})=0
\; . \qquad
\end{eqnarray}
For simplicity we do not write in Eqs. (\ref{DcW=ag+}) and
(\ref{gcUbp=0+0}) an explicit form of the terms proportional to
scalar and spinor superfields, ${\cal O}({\Upsilon})$ and $ {\cal
O}({\cal W})$. In this notation, the real part of Eq.
(\ref{gcUbp=0}) reads
\begin{eqnarray}\label{gcUbp=01}
(F(\eta -F)^{-2}){}^{ab} (\gamma_{a}\gamma_{b}{\cal W})_{q} + {\cal
O}({\Upsilon}) = 0\, . \;
\end{eqnarray}
Notice that the terms denoted here by ${\cal O}({\Upsilon})$ are
also proportional to the  fermionic fluxes (\ref{L1=def}) and vanish for the case of flat superspace. To stress this
one might use an alternative form of writing Eqs. (\ref{gcUbp=0+0})
and (\ref{gcUbp=01}),
\begin{eqnarray}\label{gcUbp=02}
(F(\eta -F)^{-2}){}^{ab} (\gamma_{a}\gamma_{b}{\cal W})_{q} + {\cal
O}({\Lambda}) = 0\, ,  \; \\
%\end{eqnarray} \begin{eqnarray}
\label{gcUbp=0+02}
(F(\delta -F)^{-2}){}^{ab} (\gamma_{a}\gamma_{b})_{pq^\prime}
D_q{\cal W}^{q^\prime} + {\cal O}({\cal W}) + {\cal O}(\Lambda)+
{\cal O}(D_q\Lambda ) =0 \; . \qquad
\end{eqnarray}

Using the worldvolume covariant derivative algebra (this is to say,
torsion and curvature describing  worldvolume geometry induced by
the superembedding, see Appendix D), one finds (from (\ref{DY=gW+}))
that
\begin{eqnarray} \label{DcW=ag+}
D_p{\cal W}^{q^\prime} &=  ia_{ab}\gamma^{ab}_{pq^\prime} +
i\tilde{a}_{ab}(\gamma^{ab}\gamma^9)_{pq^\prime}
 + {\cal O}({\Upsilon})+ {\cal O}({\cal W})\;  \qquad
\end{eqnarray}
(see Eq. (\ref{DcW=ag+1E}) of Appendix E for a more complete form).
Here $a_{ab}$ and $\tilde{a}_{ab}$ are antisymmetric tensors which
have to be determined from the further study of Bianchi identities;
note that the terms denoted by ${\cal O}({\Upsilon})+ {\cal O}({\cal
W})$ in (\ref{DcW=ag+}) do not contain irreducible parts $\propto
\gamma^{ab}_{pq^\prime}$ and $\propto
(\gamma^{ab}\gamma^9)_{pq^\prime}$. Substituting (\ref{DcW=ag+}),
one finds that  Eq. (\ref{gcUbp=0+02}) implies, in particular,
\begin{eqnarray}\label{Fxa=0}
(F(\delta -F)^{-2}){}^{ab}a_{ab} +  {\cal O}({\cal W})+ {\cal O}(
{\Lambda} )+ {\cal O}(D_p {\Lambda} )=0 \; , \qquad \\
\label{Fxta=0} (F(\delta -F)^{-2}){}^{ab}\tilde{a}_{ab} +  {\cal
O}({\cal W})+ {\cal O}({\Lambda} )+ {\cal O}(D_p {\Lambda} )=0 \; ,
\qquad
\end{eqnarray}
where we account separately for terms proportional to fermionic fluxes, ${\cal O}({\Lambda} )$, and to the Grassmann derivatives of the fermionic  fluxes, ${\cal O}(D_p{\Lambda})$ as far as these latter collect contributions from the bosonic fluxes.

The next, dim $2$ component of the Banchi identities (\ref{dG=H+R})
reads
\begin{eqnarray}\label{QdG(2)=0}
 -4i \gamma_{c\, qp}  \; (\eta -F)^{-1}{}^{ca} (G_{ab}-\tilde{q}F_{ab}) + 4i
\tilde{q}{}^\prime (h\gamma_b)_{(qp)} + 2 D_{(p}(\gamma_{b }{\cal
W})_{q)}   + && \qquad \nonumber
\\  + (\delta + i\gamma^9)_{qp}D_{b} \Upsilon + (\delta - i\gamma^9)_{qp}D_{b}
\bar{\Upsilon}  + {\cal O}({\cal W})+ {\cal O}(\Upsilon) &=& 0 \; ,
\qquad
\end{eqnarray}
where we have introduced the notation of effective charges
\begin{eqnarray}\label{tbeta:=}
 \tilde{q} := q + q^\prime \hat{C}_0  \; , \qquad
\tilde{q}{}^\prime := {q}{}^\prime e^{-\hat{\Phi}} \; \; . \qquad
\end{eqnarray}
The second term in (\ref{QdG(2)=0}) can be specified with the use of
the explicit solution (\ref{h8=}) of Eqs. (\ref{hgh=gk}),
(\ref{hg9h=}),
\begin{eqnarray}\label{sym(hg+gh=}
 & (h\gamma_b)_{(pq)}  =  - {1\over 2\sqrt{|\eta +F|}} \left[
(\gamma^{a_1a_2a_3}\gamma^9)_{pq} (\eta - F)_{b[a_1}F_{a_2a_3]} +
{1\over 4!}\gamma^{c}_{pq}\, \left((\varepsilon FFF)_{bc}  + {1\over
8}\eta_{bc} \varepsilon FFFF \right)\right]
 \; , \qquad
\end{eqnarray}
where $(\varepsilon FFF)_{ab}=\varepsilon_{abc_1\ldots c_6}
F^{c_1c_2} F^{c_3c_4} F^{c_5c_6}\;$, {\it etc}. Using Eqs.
(\ref{sym(hg+gh=}) and (\ref{DcW=ag+}), one finds that the $\propto
(\gamma^{c_1c_2c_3}\gamma^9)_{\; pq}$ irreducible part of Eq.
(\ref{QdG(2)=0}) reads, modulo terms proportional to scalar and
spinor superfields,
\begin{eqnarray}\label{1ta=1FF}   \eta_{b[c_1}
\tilde{a}_{c_2c_3]}= - {\tilde{q}^\prime \over \sqrt{|\eta
+F|}}(\eta-F)_{b[c_1} F_{c_2c_3]} \;  . \qquad
\end{eqnarray}
To be precise,  this equation is valid in its literal form in the case
of flat superspace, and modulo fermionic bilinear contributions;
 all but one  terms
in its actual {\it r.h.s.} are proportional either to the background
superspace fluxes or to the pull--back of the target space fermionic
superfields (\ref{L1=def}), and the only exception is  $ -{1\over 48
} ({\cal W}\gamma_a\gamma_{c_1c_2c_3}\gamma^9h\gamma^a\chi_b)$ (see
Eq. (\ref{1ta=1FF+OY}) in Appendix E).
 Of course, this {\it r.h.s.} also vanishes when we set to zero the additional bosonic and
fermionic superfields, $\Upsilon=0={\cal W}^q$ (which are also not
wanted from the point of view of the Q7-brane picture).

An immediate consequence of Eq. (\ref{1ta=1FF}) is  that \begin{eqnarray}\label{FF=0} F_{[bc_1}
F_{c_2c_3]}=0\; . \qquad
\end{eqnarray} Passing to a special Lorentz frame where $F_{ab}$
has Darboux's standard form, $F_{ab}=
i\sigma^2\otimes diag (f_1  ,f_2 , f_3 , f_4)$, one easily finds that, for any
solution of Eq. (\ref{1ta=1FF}), only one of four 'eigenvalues'
$f_1, f_2, f_3, f_4$ might be nonzero. In other words, the general
solution of $F_{[bc_1} F_{c_2c_3]}=0$ is given by
$F_{ab}=2k_{[a}l_{b]}\, f$ with $l^2=l_al^a=-1$, $k_ak^a=\pm 1$ and
an arbitrary function $f$. Then $F_{b[c_1} F_{c_2c_3]}=0$ and,
hence, $\tilde{a}_{ab}= - {\tilde{q}^\prime \over
\sqrt{|\eta+F|}}F_{ab}$. To resume,
\begin{eqnarray}\label{Fab=klf}
 F_{ab}=2k_{[a}l_{b]}\, f \; , \qquad l^2=l_al^a=-1\;  , \qquad k^2=k_ak^a=\pm 1\;  , \qquad \\
 \label{ta=q1F} \tilde{a}_{ab}= - {\tilde{q}^\prime \over \sqrt{|det (\eta+F)|}}\; F_{ab}= - {
 \tilde{q}^\prime \, 2k_{[a}l_{b]}f\over \sqrt{|1-f^2|}}
 \; . \qquad
\end{eqnarray}
Now, using the consequence (\ref{Fxta=0}) of Eq. (\ref{gcUbp=0+02}),
which on the level of accuracy of our present discussion reads
$(F(\delta -F)^{-2}){}^{ab}\tilde{a}_{ab}=0$, one concludes that, for $\tilde{q}^\prime\not= 0$, 
\begin{eqnarray}\label{f2=0}
f^2=0
\; . \qquad
\end{eqnarray}
Thus the system of Eq. (\ref{1ta=1FF}) and (\ref{gcUbp=0+02}) has
only trivial solutions, so that  $\tilde{a}_{ab}=0$ and
\begin{eqnarray}\label{q'Fab=0}
\tilde{q}^\prime F_{ab}=0 \;  . \qquad
\end{eqnarray}

Of course, one might notice that our discussion at this point  is rough enough as far as, in particular, in our superspace context, Eq.  (\ref{f2=0}) implies just the nilpotence of $f$ rather than $f=0$, so that Eq. (\ref{q'Fab=0}) should be rather replaced by the statement of vanishing of the 'pure bosonic body' (not nilpontent part)  of $F_{ab}$. Next, even ignoring this simplest nilpotent contribution, one might note that, in the case of nonvanishing fermionic (super)fields and fluxes we should rather have $\tilde{q}^\prime F_{ab}={\cal O}(\Upsilon) + {\cal O}({\cal W})$.
But  such a {\it r.h.s.}  (either from the
additional scalar and spinor fields $\Upsilon$ and ${\cal W}$, which do  not
have good treatment in discussions of hypothetical Q7-branes, or from the nilpotent $f$ in (\ref{Fab=klf}))  does not prevent from the
conclusion that the selfconsistency does not allow for a second
(actually first) dynamical gauge field and that for $q^\prime\not=
0$ {\it our constrains rather describe a fixed field configuration
than a dynamical system with two gauge field potentials}.

\subsection{Conclusion}

Thus we have shown that, imposing the type IIB 7--brane
superembedding equation together with two different sets of gauge
field constraints, requires, instead of dynamical equations, that
one of the field strengths vanish, $F_{ab}=0$ (or expressed through
the additional scalar and fermionic fields introduced artificially
to complete the other field strength till a supermultiplet). This
shows that there is no place for Q7--brane in the standard setup of
the superembedding approach.

In particular, such a superembedding description, leading to  Eq. (\ref{q'Fab=0}) or to $\tilde{q}^\prime F_{ab}={\cal O}(\Upsilon) + {\cal O}({\cal W})$, cannot be used to construct an action by superembedding based method of \cite{Howe-GAP}, starting from the candidate Wess--Zumino term of Eqs. (\ref{LWZQ7=QL}), (\ref{LWZRS=}). Hence, this procedure which applies for all known branes, fails in the case of hypothetical Q7--brane. This suggests to conclude that a supersymmetric and $\kappa$-symmetric Q7-brane action does not exist.

One can also solve (\ref{q'Fab=0}) by setting $q^\prime=0$, but this
would imply that two gauge fields are of the same type and actually,
when $\Upsilon=0={\cal W}$, would result in their coincidence. This
does not correspond to a Q7-brane, as far as the corresponding
charge matrix Q is degenerate, but rather to a (S)D7--brane itself
(see Eqs. (\ref{D7-Q7}) and (\ref{Q7=q+q})).

\section{Summary, Conclusions and Discussion. \label{Conclusion}}

The main aim of our present study was to search for possible
description of hypothetical Q7--branes dynamics  in the frame of
superembedding approach.

We begin by the  analysis of candidate Wess--Zumino (WZ) term which
confirmed the suggestion of previous studies in purely bosonic limit
\cite{Dima+Eric07} on that the Q7-brane, if existed, would carry out
two worldvolume gauge fields on its worldvolume as well as on the
general structure of the candidate WZ term constructed
with these two gauge fields. This could be used to construct the
complete action in an algorithmic manner if one lifts this
WZ term to a 9--form in the 7-brane worldvolume superspace
with 8 bosonic and 16 fermionic `directions', $W^{(8|16)}$. This, in
its turn, would be possible if it existed the possibility to
describe two different gauge potentials by superfields on this
worldvolume superspace.

Motivated by this observation, we studied the possibilities to
describe two independent dynamical gauge fields on the worldvolume
of a type IIB 7--brane. If consistent, this would be the description
of the Q7--brane of \cite{Dima+Eric07}. However, the result of our
study is negative.

The 7-brane superembedding equation is off-shell. Imposing the
constraints on the generalized field strength
$F_2=dA-q\hat{B}_2-q^\prime\hat{C}_2$ of the worldvolume gauge
superfield (super-1-form $A$) leads to the equations of motion for
SD7-brane related to D7-brane by SL(2) transformation (called '(p,q)
7-branes' in \cite{Dima+Eric07}). The case of D7--brane itself
corresponds to $(q,q^\prime )= (1,0)$, and, in the light of
existence of the SL(2) covariant formalism \cite{DpSL2}, it is
sufficient to use this charge configuration as a reference point.

Then the natural hypothesis, also supported by the results on the
bosonic Q7-brane action and inclusion of fermions in the linear
approximation, is that Q7--brane might be described by a SD7-brane
carrying an additional gauge field on its worldvolume. If this were
the case, imposing the superspace constraints of an additional  SYM
super-1-form leaving on a D7-brane worldvolume we would obtain a
consistent system of dynamical equations which would be the
equations of motion for the Q7-brane.

We have done this in Sec. 4, and the result of this analysis has
been negative. Namely, the consistency of the second gauge field
constraints requires that either these are of the same type as the
original one, and then the second field strength coincide with the
first one, $G_{ab}=F_{ab}$, or, for the different second field
strength (including different combination of RR and NS-NS fields in
its definition), it results in $F_{ab}=0$. This relation  can
characterize a field configuration rather than equations of motion
of a dynamical brane with two gauge fields.

This suggests that, probably, Q7-brane does not exists as
supersymmetric dynamical system,  but there exists only the Q7-brane
BPS state described by a particular solution of supergravity
equations \cite{Eric+06sD7,Eric+06aD7} and also of some
non-supersymmetric system of worldvolume equations. If this is the
case, then, when imposing the requirement of supersymmetry (as we do
when develop superembedding approach, by its construction
\cite{bpstv}--\cite{Howe+Sezgin04}) this supersymmetric solution
describes the only field configuration which is allowed by
consistency of the system of manifestly supersymmetric equations,
thus resulting in the solution rather than in a system of dynamical
equations (as it would be the case if we were considering a true
dynamical superbrane). Actually, a search for particular BPS
solution of such a type might be an interesting new application of
the superembedding approach.

\bigskip

One might also think that more general setup is needed to conclude
definitely about non-existence of Q7--brane. First idea might be to
try to introduce, instead of the second gauge field, a 5--form potential the
generalized field strength of which, ${\cal F}_{abcdef}$, would be
dual to the one of the second gauge field strength, $G_{ab}$, on the
mass shell. However, one can easily check, using  the natural form
of the $\tilde{{\cal F}}_6$ superspace constraints
\begin{eqnarray}
\label{tF6=} \tilde{{\cal F}}_6:= dA_5 - \hat{C}_6 - {\cal {F}}_2
\wedge \hat{C}_4 - {1\over 2} {\cal {F}}_2 \wedge {\cal {F}}_2
\wedge {{C}}_2 = {1\over 6!}e^{a_6} \wedge \ldots \wedge e^{a_1}
F_{a_1\ldots a_6} \; ,   \qquad
\end{eqnarray}
that, when considered on the D7--brane worldvolume superspace, the
Bianchi identities for this constraint imply a kind of {\it duality
to the 1-form gauge field strength entering the superembedding
supermultiplet},
\begin{eqnarray}\label{F6=*F2+}
 e^{\hat{\Phi}} \sqrt{|\eta + F|} (\eta-F)^{-1}{}_{a}{}^{b}  \; F_{bb_1\ldots b_5}
 = {1\over 2 }\varepsilon_{ab_1\ldots b_5cd}F^{cd} + {\cal O}(FF) \; , \qquad
\end{eqnarray}
and not to some second independent gauge fields strength, as wanted.

One might also think on that our constraints for the two gauge
fields on 7-brane worldvolume, (\ref{F2=dA-B2-2}) and
(\ref{dA-B-C=}), are too strong, and search for more general ones.
However, we notice that i) already the inclusion of additional
superfields $\Upsilon$ and ${\cal W}^q$ in (\ref{dA-B-C=}) creates a
problem as these do not have any place in the Q7-brane picture as
conjectured in \cite{Dima+Eric07}, and also that ii) these
constraints are consistent on linearized level leading to a correct
free field equations.

Then one might think that a possibility to describe Q7--brane might
appear when one rejects imposing the superembedding equation.

Although our study does not give a definite answer on this query,
and so a possibility of radical reformulation of superembedding
approach to incorporate new exotic superbranes is not ruled out,  we
do not expect this to be the case for Q7-branes\footnote{A
modification of superembedding equation does occur in the 'minus one
quantization' picture of \cite{Howe+Linstrom+Linus}, where the
non-Alebian structure of coincident D$p$--brane is described by the
boundary fermions. But in this case the boundary fermions are
included in the set of worldvolume superspace coordinates, but not in the set of
tangent superspace ones, so that, similarly to superfield
description of spinning string, the number of fermionic dimensions
of the worldsheet superspace exceeds the one of the  target space.
This is not the case for superembedding description of 1/2 BPS
superbranes, where the target superspace has twice more fermionic
dimensions that the worldvolume superspace.}. The reasons are that
all the known dynamical branes ($1/2$ BPS superbranes) allow for the
description by worldvolume superspace whose embedding is
characterized by the superembedding equation (either alone, or together with
other constraints), and also that in the case of 7-branes the
superembedding equation is off-shell, {\it i.e.} it is not strong
enough even to produce equations of motion.

Finally, one could propose to assume that Q7--brane might be a brane
which cannot be described by the superembedding approach. But,
again, no examples of such a situation are known and no reason are
seen for a dynamical brane to do not allow for an on-shell
superfield superembedding description.

\bigskip

The above arguments suggest (although, of course, do not prove) the
conjecture that Q7--brane of \cite{Eric+06sD7,Dima+Eric07} is not a
dynamical brane, but just  a particular BPS configuration of the
system of two different SD$7$--branes; Q7--branes can be described by
the supersymmetric solution \cite{Eric+06sD7} of type IIB
supergravity equations, but neither $\kappa$--symmetric and
supersymmetric action, nor supersymmetric equations of motion may be
associated to them. If this is the case, the superembedding
description of such a BPS state is also given by a configuration of
worldvolume superfields, rather than by (super)fields obeying
dynamical equations of motion, and this is exactly what we have
observed in our study of the superembedding approach to 7-branes.

\bigskip

Our study and the above discussion might provide a suggestion for
the old standing problem of that the commonly accepted action for
coincident D$p$--branes \cite{Myers:1999ps} does not possess neither
supersymmetry nor Lorentz symmetry (see footnote \ref{footnote5} for
references on recent progress). The situation with Q7--brane seems
to be similar: one can write down a bosonic action
\cite{Dima+Eric07} (and in this case, in contrast to
\cite{Myers:1999ps}, at least the Lorentz SO(1,9) symmetry is
respected), but our study suggests that supersymmetric
generalization of the equations of motion fails and so would fail an
attempt of supersymmetric generalization of that action. It may be
that in both cases  we deal with  an action for a dynamical system
of several branes (two different SD7-brane in Q7-case or N
D$p$-branes in the 'dielectric brane' action of \cite{Myers:1999ps})
which is not supersymmetric.

How it could be if, speaking about multi-brane system, one implies
an action obtained in some way from a sum of two or more
SD$7$--brane actions, which are supersymmetric by construction? The
possible answer is that  the sum of, say, two SD$7$--brane actions
{\it lost} supersymmetry when one requires the existence of an
intersection or, more generally, of some set of common points of two
worldvolumes (see \cite{BK99}).

To clarify this issue, it seems convenient to speak about
intersections and in terms of local supersymmetry preserved by
bosonic brane actions \cite{BdAI1,IB+JdA05} which represents the
$\kappa$--symmetry of the original superbranes. On the intersection
(or on a set of common points) we, roughly speaking, should impose
on the local supersymmetry parameter the condition that it vanishes
by the two projectors corresponding to the $\kappa$--symmetries of
two intersecting branes. But this system of two equation have
nontrivial solutions only for definite angles of the intersection
\cite{kappaInt}. On the other hand, considering the action principle
or effective equations of motion for a bound state (for a composed
system), it does not look proper to fix the angle of the
intersection. Thus we have an action of, say, string ending on a
D$p$--brane in which the supersymmetry is broken on the
intersection; however, there exists a ground state solution where
this broken supersymmetry  is restored \cite{BK99}\footnote{In
\cite{Cederwall:1997hg} an introduction of additional degrees of
freedom on the intersection was proposed to take care of this
'anomaly'. Such additional degrees of freedom might come  from
supergravity part of the superbrane--supergravity dynamical system
(see \cite{IB+JdA05} and refs therein for the complete but gauge
fixed Lagrangian description of such a system). Such a point of view
seems to be in agreement with resent description  of coincident
D-branes by boundary fermions at the ends of string
\cite{Howe+Linstrom+Linus}. Indeed, to reproduce the non-Abelian
structure of the coincident D$p$-brane actions form the approach of
\cite{Howe+Linstrom+Linus}, the quantization of boundary
fermions is to be done. In such a quantization  the Myers action is reproduced
\cite{Howe+Linstrom+Linus} and the Lorentz covariance is lost. To
obtain the completely covariant and supersymmetric  result one
(presumably) needs to carry out the complete quantization of the
string model with boundary fermions, and not just the boundary
fermion sector. However, the complete quantization of open
superstring should also reproduce the supergravity fields (from the
closed string sector). Their influence might then come in the form
of the above mentioned additional degrees of freedom in the
effective low energy multiple brane action. The further analysis of
these issues goes beyond the score of this paper. }.

Now, in the superembedding approach we are searching for
supersymmetric field configurations (usually, for the supersymmetric
equations of motion). What we have to find for a non-supersymmetric
system with a particular supersymmetric solution? - All the
conditions for the fields which guarantee that the field
configuration is supersymmetric. Thus we have to arrive at a
particular supersymmetric field configuration rather than at a set
of supersymmetric equations. This is just what happens in the case
of Q7-brane as described by (S)D7-brane carrying an additional
worldvolume gauge field: the preservation of 1/2 of the
supersymmetry in the presence of the second, different gauge field
implies vanishing of the 'original' (S)D7-brane gauge fields (modulo
scalar fields and fermions, but this does not change the
conclusion);  this is characteristic for a solution rather than for
dynamical equations describing a dynamical object, and this is just
the solution which possesses supersymmetry.

Of course, it is always hard to prove non-existence of some object,
and our results cannot be considered as a proof. However, on one
hand  we find the above arguments against the existence of a
Q7-brane as a dynamical supersymmetric object convincing and,  on
the other hand, we hope that, if it finally were found that some
exotic possibility for constructing the supersymmetric action and/or
equations of motion for Q$7$--branes happened to exist, our
discussion in this paper would be helpful in search for such an
exotic construction, e.g. for an exotic modification of the
superembedding approach.

\bigskip

\section*{Acknowledgments}

The author thanks Paolo Pasti, Mario Tonin and especially Dima
Sorokin and Jos\'e~de~Azc\'arraga for discussions on different
stages of this work which has been partially supported by research
grants from the Spanish MCINN (FIS2008-1980), the INTAS (2006-7928),
and the Ukrainian National Academy of Sciences and Russian RFFI
grant 38/50--2008.

\bigskip

{\small

\setcounter{section}{0}
\renewcommand{\thesection}{}
\section{Appendix A: Spinors in ten and eight dimensions}
\renewcommand{\theequation}{A.\arabic{equation}}
\setcounter{equation}{0}
\renewcommand{\thesection}{A}
\setcounter{subsection}{0}

In $D=10$ we use the Majorana--Weyl spinors and real symmetric sigma
matrices, 
$\sigma^{\underline{a}}_{\alpha\beta}=\sigma^{\underline{a}}_{\beta\alpha}$
and
$\tilde{\sigma}_{\underline{a}}^{\alpha\beta}=\tilde{\sigma}_{\underline{a}}^{\beta\alpha}$, 
which obey \begin{eqnarray} \label{SO(1,9)sigmas}
\sigma^{\underline{a}}\tilde{\sigma}^{\underline{b}}+
\sigma^{\underline{b}}\tilde{\sigma}^{\underline{a}}=\eta^{\underline{a}\underline{b}}\;
, \qquad && \tilde{\sigma}^{\underline{a}}{\sigma}^{\underline{b}}+
\tilde{\sigma}^{\underline{b}}
{\sigma}^{\underline{a}}=\eta^{\underline{a} \underline{b}}\; ,
\qquad \\ \nonumber &&  \eta^{\underline{a}\underline{b}}= diag
({1,-1,...,-1})\; , \qquad {\underline{a}\, ,
\underline{b}}=0,1,\ldots , 9\; , \qquad
\end{eqnarray}
as well as  the famous $D=10$ identity \begin{eqnarray}
\label{10Did=}\sigma_{\underline{a}\,
(\alpha\beta}\sigma^{\underline{a}}{}_{\gamma )\delta}=0 \; , \qquad
\tilde{\sigma}^{\underline{a}\,
(\alpha\beta}\tilde{\sigma}_{\underline{a}}{}^{\gamma )\delta}=0 \;
. \qquad
\end{eqnarray}
We define
$\sigma^{\underline{a}\underline{b}}=\sigma^{[\underline{a}}\tilde{\sigma}^{\underline{b}]}=
{1 \over
2}(\sigma^{\underline{a}}\tilde{\sigma}^{\underline{b}}-\sigma^{\underline{a}}\tilde{\sigma}^{\underline{b}})$
as well as
$\tilde{\sigma}^{\underline{a}\underline{b}}=\tilde{\sigma}^{[\underline{a}}{\sigma}^{\underline{b}]}$,
$\sigma^{\underline{a}\underline{b}\underline{c}}=\sigma^{[\underline{a}}\tilde{\sigma}^{\underline{b}}\sigma^{\underline{c}]}$
etc. Matrices ${\sigma}_{\underline{a}}$,
${\sigma}_{\underline{a}_1\ldots \underline{a}_5}$ and
$\tilde{\sigma}_{\underline{a}}$,
$\tilde{\sigma}_{\underline{a}_1\ldots \underline{a}_5}$ are
symmetric, ${\sigma}_{\underline{a}_1\underline{a}_2
\underline{a}_3}$ and
$\tilde{\sigma}_{\underline{a}_1\underline{a}_2\underline{a}_3}$ are
antisymmetric, furthermore, ${\sigma}_{\underline{a}_1\ldots
\underline{a}_5}$ is self-dual while
$\tilde{\sigma}_{\underline{a}_1\ldots \underline{a}_5}$ is
anti-self dual.

In eight dimensions, $d=1+7$, one can define 16-component
pseudo-real (pseudo-Majorana) spinors
\begin{eqnarray}
\label{SO(1,7)spinor} (\chi^q)^* = \gamma^0_{qp}\chi^p\; , \qquad
q,p=1, \ldots , 16
 \; .
\end{eqnarray}
The charge conjugation matrix is symmetric and can be identified
with the the unity matrix $\delta_{qp}$, the seven gamma matrices
$\gamma^a_{qp}$ are symmetric and pseudo real,
\begin{eqnarray}
\label{SO(1,7)gammas} \gamma^a_{qp}= \gamma^a_{pq}\; , \qquad
(\gamma^a)^* = \gamma^0 \gamma^a \gamma^0 \; ,  \qquad a=0\, ,\, 1
\, , \, \ldots \, ,\, 7 \; ,  \qquad q,p=1, \ldots , 16
 \; .
\end{eqnarray}
This is also the case for their product which we call
$\gamma^9_{qp}$,
\begin{eqnarray}
\label{gamma9} \gamma^9_{qp}:= \gamma^0 \gamma^1\ldots \gamma^7 =
\gamma^9_{pq}\; , \qquad (\gamma^9)^* = \gamma^0 \gamma^9 \gamma^0 =
- \gamma^9 \; .  \qquad
\end{eqnarray}
This can be used  to construct two chiral projectors ${1\over
2}(\delta\pm i\gamma^9)$
\begin{eqnarray}
\label{1+ig9} \delta_{qp}=
 {1\over 2}(\delta +  i\gamma^9)_{qp}+ {1\over 2}(\delta -  i\gamma^9)_{qp}\; , \qquad
\cases{{1\over 2}(\delta +  i\gamma^9){1\over 2}(\delta +
i\gamma^9)= {1\over 2}(\delta + i\gamma^9)\; ,  \cr {1\over
2}(\delta -  i\gamma^9){1\over 2}(\delta - i\gamma^9)= {1\over
2}(\delta - i\gamma^9) \cr {1\over 2}(\delta +  i\gamma^9){1\over
2}(\delta -  i\gamma^9)= 0 }\; ,  \qquad \\ \nonumber (\delta +
i\gamma^9)^* = (\delta +  i\gamma^9)= \gamma^0 (\delta -
i\gamma^9)\gamma^0 \; . \qquad
\end{eqnarray}

Let us notice the $d=8$ gamma--metrix identity
\begin{eqnarray}
\label{d8Iden} \gamma_{b\; (q_1q_2}\, \gamma^b_{q_3 )p} &=&
\delta_{(q_1q_2}\, \delta_{q_3 )p} + \gamma^9_{(q_1q_2}\,
\gamma^9_{q_3 )p}= \qquad \nonumber \\ &=& {1\over 2} (\delta-
i\gamma^9)_{(q_1q_2} (\delta + i\gamma^9)_{q_3 )p} + {1\over 2}
 (\delta+ i\gamma^9)_{(q_1q_2} (\delta - i\gamma^9)_{q_3 )p} \; .
 \qquad
\end{eqnarray}
which can be derived {\it e.g.} from the $D=10$ identity
(\ref{10Did=}). Useful consequences of (\ref{d8Iden}) are
\begin{eqnarray}
\label{p+gxp+g=p+xp-} ((\delta + i\gamma^9)\gamma_{b})_{(q_1| q}\, ((\delta + i\gamma^9)\gamma^b)_{|q_2)p} = (\delta + i\gamma^9)_{q_1q_2}\, (\delta - i\gamma^9)_{qp} \; , \qquad \\ \label{p-gxp-g=p-xp+} ((\delta - i\gamma^9)\gamma_{b})_{(q_1 |q}\, ((\delta - i\gamma^9)\gamma^b)_{|q_2)p} = (\delta - i\gamma^9)_{q_1q_2}\, (\delta + i\gamma^9)_{qp}
\; .
 \qquad
\end{eqnarray}

As it is well known, in $SO(1,9)$  covariant notation ($D=10$) the
basis of symmetric $16\times 16$ matrices is provided by $10$
$\sigma^{\underline{a}}_{\alpha\beta}$ and $126$ (five index
self--dual) $\sigma^{\underline{a}_1\ldots
\underline{a}_5}_{\alpha\beta}:= {1\over
5!}\varepsilon^{\underline{a}_1\ldots
\underline{a}_5\underline{b}_1\ldots \underline{b}_5}
\sigma_{\underline{b}_1\ldots \underline{b}_5\alpha\beta}$, while
the basis of the antisymmetric matrices is provided by $120$
$\sigma^{\underline{a}_1\ldots \underline{a}_3}_{\alpha\beta}$. In
the $SO(1,7)$  covariant notation ($d=8$) this corresponds to the
following bases of the symmetric and antisymmetric matrices:
\begin{eqnarray}
\label{SYMbasis}  SYMM\; : & \; \delta_{qp}\, , \; \gamma^9_{qp}\, ,
\; \gamma^a_{qp}\, , \; (\gamma^{abc}\gamma^9)_{qp}\, , \;
(\gamma^{abcd})_{qp}\, , \qquad
1+1+8+56+70=136 \; , \qquad \\
\label{aSYMbasis}  AntiSYMM\; : & \;  (\gamma^{a}\gamma^9)_{qp}\, ,
\; \gamma^{ab}_{qp}\, , \; (\gamma^{ab}\gamma^9)_{qp}\, , \;
(\gamma^{abc})_{qp}\, , \qquad 8+28+28+56=120 \; .
 \qquad
\end{eqnarray}
Notice that $(\gamma^{abcd}\gamma^9)_{qp}= {1\over
4!}\varepsilon^{abcda^\prime b^\prime c^\prime d^\prime}
(\gamma_{a^\prime b^\prime c^\prime d^\prime})_{pq}$ and, hence, is
not independent.

\setcounter{section}{0}
\renewcommand{\thesection}{}
\section{Appendix B. Type IIB supergravity constraints and their consequences}
\renewcommand{\theequation}{B.\arabic{equation}}
\setcounter{equation}{0}
\renewcommand{\thesection}{B}
\setcounter{subsection}{0}

Denoting the fermionic vielbein of type IIB superspace as in
(\ref{Eua-cE}),
\begin{eqnarray}\label{cE=E1E2}
 {\cal E}^{{\underline{\beta}}}:= {\cal E}^{{{\beta}i}}:= (E^{\beta 1}\, , \; E^{\beta 2}) \; ,
\end{eqnarray}
one can write the bosonic torsion constraint in the form of
(\ref{Tua=}),
\begin{eqnarray}
\label{Tub=II} {T}^{\underline{a}} := D{E}^{\underline{a}}=  -i
{\cal E}\wedge \underline{\sigma}^{\underline{a}} {\cal E}\; ,
\qquad
\underline{\sigma}^{\underline{a}}_{{\underline{\alpha}}{\underline{\beta}}}
:= \sigma^{\underline{a}}_{\alpha\beta} \delta_{ij}\; .
\end{eqnarray}

The fermionic torsion forms of type IIB superspace
$T^{{{\underline{\beta}}}}:= D {\cal E}^{{\underline{\beta}}}=
(T^{\beta 1}\, , \; T^{\beta 2})$ are
\begin{eqnarray}\label{Tal1=str}
 T^{\alpha 1} &=& -E^{\alpha 1}\wedge E^{\beta 1} \nabla_{\beta 1}e^{-{\Phi}} + {1\over
2} E^{1}\sigma^{\underline{a}}\wedge E^{1}\,
\tilde{\sigma}_{\underline{a}}^{\alpha\beta} \nabla_{\beta
1}e^{-{\Phi}} + \qquad \nonumber \\ && \qquad +
E^{\underline{a}}\wedge {\cal E}^{{\underline{\beta}}}
T_{{\underline{\beta}}\underline{a}}{}^{\alpha 1 }
 +  {1\over 2}E^{\underline{b}} \wedge E^{\underline{a}}
 T_{\underline{a}\underline{b}}{}^{\alpha 1 }  \; ,  \qquad \\
\label{Tal2=str}  T^{\alpha 2} &=& -E^{\alpha 2}\wedge E^{\beta 2}
\nabla_{\beta 2}e^{-{\Phi}} + {1\over 2}
E^{2}\sigma^{\underline{a}}\wedge E^{2}\,
\tilde{\sigma}_{\underline{a}}^{\alpha\beta} \nabla_{\beta
2}e^{-{\Phi}} + \qquad \nonumber \\ && \qquad +
E^{\underline{a}}\wedge {\cal E}^{{\underline{\beta}}}
T_{{\underline{\beta}}\underline{a}}{}^{\alpha 2 }
 +  {1\over 2}E^{\underline{b}} \wedge E^{\underline{a}}
 T_{\underline{a}\underline{b}}{}^{\alpha 2 }  \; , \qquad
\end{eqnarray}
where
\begin{eqnarray}\label{Tfbf=}
T_{{\underline{\alpha}}\underline{b}}{}^{{\underline{\gamma}}} &=:&
- t_{\underline{b}{\underline{\alpha}}}{}^{{\underline{\gamma}}} = -
{1\over 8} H_{\underline{b}\underline{c}\underline{d}}\,
(\sigma^{\underline{c}\underline{d}}\tau_3)_{{\underline{\alpha}}}{}^{{\underline{\gamma}}}
+ \sum\limits_{n=0}^{4} (\sigma_{\underline{b}}
\tilde{R}\!\!\!/{}^{(2n+1)}\tau_1(\tau_3)^n)_{{\underline{\alpha}}}{}^{{\underline{\gamma}}}
= \qquad \\ \nonumber &=&  - {1\over 8} \left(
H_{\underline{b}\underline{c}\underline{d}}\,
\sigma^{\underline{c}\underline{d}}\tau_3 - \sigma_{\underline{b}}
\tilde{R}\!\!\!/{}^{(1)}i\tau_2 + \sigma_{\underline{b}}
\tilde{R}\!\!\!/{}^{(3)}\tau_1 - -\!\!\!\! ^1_2 \;
\sigma_{\underline{b}} \tilde{R}\!\!\!/{}^{(5)}i\tau_2
\right){}_{{\underline{\alpha}}}{}^{{\underline{\gamma}}}\; , \qquad
\end{eqnarray}
\begin{eqnarray}\label{R/=}
\tilde{R}\!\!\!/{}^{(2n+1)}:= {1\over (2n+1)!}
R_{\underline{a}_1\ldots \underline{a}_{2n+1}}
\tilde{\sigma}^{\underline{a}_1\ldots \underline{a}_{2n+1}} \; .
\qquad
\end{eqnarray}

Other useful equations are
\begin{eqnarray}\label{DDPhi=}
D_{\hat{\underline{\beta}}}D_{\hat{\underline{\gamma}}} e^{-{\Phi}}
= i
\sigma^{\underline{a}}_{\hat{\underline{\beta}}\hat{\underline{\gamma}}}
D_{\underline{a}} e^{-{\Phi}}  - i
(\tilde{R}\!\!\!/{}^{(1)}i\tau_2)_{\hat{\underline{\beta}}\hat{\underline{\gamma}}}
+ {i\over 2}
(\tilde{R}\!\!\!/{}^{(3)}\tau_1)_{\hat{\underline{\beta}}\hat{\underline{\gamma}}}
+ {i\over 2}
(\tilde{H}\!\!\!/{}^{(3)}\tau_3)_{\hat{\underline{\beta}}\hat{\underline{\gamma}}}
\; . \qquad
\end{eqnarray}
and
\begin{eqnarray}\label{Tfff=str}
& T_{{\underline{\beta}}
{\underline{\gamma}}}{}^{\underline{\alpha}} D_{\underline{\alpha}}=
\left(\matrix{ -2 \left(\delta_{(\beta }{}^\alpha  D_{\gamma )1}
e^{-{\Phi}}  - {1\over 2}
\sigma^{\underline{a}}_{\beta\gamma}\tilde{\sigma}_{\underline{a}}^{\alpha\delta}
D_{\delta 1}e^{-{\Phi}}\right)\,  D_{\alpha 1}   \hspace{1.5cm} 0
\hspace{1.5cm} \cr \hspace{1.5cm} 0 \hspace{1.5cm} -2
\left(\delta_{(\beta }{}^\alpha  D_{\gamma )2} e^{-{\Phi}} - {1\over
2}
\sigma^{\underline{a}}_{\beta\gamma}\tilde{\sigma}_{\underline{a}}^{\alpha\delta}
\, D_{\delta 2}e^{-{\Phi}}\right)\,  D_{\alpha 2}  } \right)\; .
\quad
\end{eqnarray}

The constraints for the NS--NS three form field strength are
\begin{eqnarray}
\label{H3=IIBstr} && H_{3} = - i {E}^{\underline{a}}\wedge
({E}^{1}\wedge \sigma_{\underline{a}} {E}^{1} -  {E}^{2}\wedge
\sigma_{\underline{a}} {E}^{2}) +{1\over 3!}
{E}^{\underline{c}_3}\wedge{E}^{\underline{c}_2}\wedge
{E}^{\underline{c}_1}
H_{\underline{c}_1\underline{c}_2\underline{c}_3}\; . \qquad
\end{eqnarray}
The RR field strengths obey the constraints
\begin{eqnarray}\label{R=IIBstr}
 R_{2n+1} &=&  2i e^{-{\Phi}} E^{\alpha 2} \wedge E^{\beta 1}
 \wedge \bar{\sigma}^{(2n-1)}_{\alpha\beta} -
  e^{-{\Phi}} \left(E^2\wedge \bar{\sigma}^{(2n)}\nabla_1\Phi
- (-)^n E^1\wedge \bar{\sigma}^{(2n)}\nabla_2\Phi\right) +  \qquad
\nonumber \\ &+& {1 \over (2n+1)!} E^{\underline{a}_{2n+1}} \wedge
... \wedge E^{\underline{a}_1} R_{\underline{a}_1\ldots
\;\underline{a}_{2n+1}}\; , \quad
\end{eqnarray}
where \begin{eqnarray}\label{barsigma=} &
\bar{\sigma}^{(2n+1)}_{\alpha\beta} := {1 \over (2n+1)!}
E^{\underline{a}_{2n+1}} \wedge ... \wedge E^{\underline{a}_1}
(\sigma_{\underline{a}_1\ldots
\;\underline{a}_{2n+1}})_{\alpha\beta}\;  . \qquad
\end{eqnarray}
Eq. (\ref{R=IIBstr}) includes, as particular cases,
\begin{eqnarray}\label{R1=str}
&  R_1= E^2 \,\nabla_1e^{-{\Phi}} - E^1\,\nabla_2e^{-{\Phi}}
   + {E}^{\underline{c}}
R_{\underline{c}}\; ,  \qquad  \\
\label{R3=str} & R_{3} = 2ie^{-{\Phi}} E^2 \wedge E^1 \wedge
\sigma^{(1)} + E^2 \wedge \sigma^{(2)}\,\nabla_1e^{-{\Phi}} + E^1
\wedge \sigma^{(2)}\,\nabla_2e^{-{\Phi}}
  + \qquad \nonumber \\ & +{1\over
3!} {E}^{\underline{c}_3}\wedge{E}^{\underline{c}_2}\wedge
{E}^{\underline{c}_1}
R_{\underline{c}_1\underline{c}_2\underline{c}_3}\; , \qquad \\
\label{R5=str}
 R_{5} &= 2i  e^{-{\Phi}}E^{\alpha 2} \wedge E^{\beta 1}
 \wedge \bar{\sigma}^{(3)}_{\alpha\beta} +
  \left(E^2\wedge \bar{\sigma}^{(4)}\nabla_1e^{-{\Phi}}
-  E^1\wedge \bar{\sigma}^{(4)}\nabla_2e^{-{\Phi}}\right) +  \qquad
\nonumber \\ &+ {1 \over 5!} E^{\underline{a}_{5}} \wedge ... \wedge
E^{\underline{a}_1} R_{\underline{a}_1\ldots \;\underline{a}_{5}}\;
,
\end{eqnarray}
as well as the constraints for the super--field strength of the
higher forms
\begin{eqnarray}\label{R7=str}
 R_{7} &=&  2i e^{-{\Phi}} E^{\alpha 2} \wedge E^{\beta 1}
 \wedge \bar{\sigma}^{(5)}_{\alpha\beta} -
  e^{-{\Phi}} \left(E^2\wedge \bar{\sigma}^{(6)}\nabla_1\Phi
+ E^1\wedge \bar{\sigma}^{(6)}\nabla_2\Phi\right) +  \qquad
\nonumber \\ &+& {1 \over 7!} E^{\underline{a}_{7}} \wedge ...
\wedge E^{\underline{a}_1} R_{\underline{a}_1\ldots
\;\underline{a}_{7}}\; ,
\\
\label{R9=str}
 R_{9} &=&  2i e^{-{\Phi}} E^{\alpha 2} \wedge E^{\beta 1}
 \wedge \bar{\sigma}^{(7)}_{\alpha\beta} -
  e^{-{\Phi}} \left(E^2\wedge \bar{\sigma}^{(8)}\nabla_1\Phi
- (-)^n E^1\wedge \bar{\sigma}^{(8)}\nabla_2\Phi\right) +  \qquad
\nonumber \\ &+& {1 \over 9!} E^{\underline{a}_{9}} \wedge ...
\wedge E^{\underline{a}_1} R_{\underline{a}_1\ldots
\;\underline{a}_{9}}\; ,
\end{eqnarray}
whose higher-dimensional purely bosonic tensor parts are dual to the
bosonic field strengths of the lower forms
\begin{eqnarray}\label{Rq=*RD-str}
& R_{\underline{a}_1 \ldots \underline{a}_{9-2n}} = {(-)^n \over
(2n+1)!}\; \varepsilon_{\; \underline{a}_1 \ldots
\underline{a}_{9-2n}\underline{b}_1 \ldots \underline{b}_{2n+1}}
R^{\underline{b}_1 \ldots \underline{b}_{2n+1}} \; \qquad  :  \\
\label{R9=*R1str} & R_{\underline{a}_1 \ldots \underline{a}_{9}} =
\,  \varepsilon_{\; \underline{a}_1 \ldots
\underline{a}_{9}\underline{b}} R^{\underline{b}} \; , \qquad R_{a_1
\ldots a_{7}} = - {1\over 3!} \varepsilon_{\; a_1 \ldots a_{7}b_1
\ldots b_{3}} R^{b_1 \ldots b_{3}} \; , \qquad
\\
\label{R5=*R5str} & R_{a_1 \ldots a_5} = (\ast R)_{a_1 \ldots a_5}
:= {1\over 5!} \varepsilon_{\; a_1 \ldots a_5b_1 \ldots b_5} R^{b_1
\ldots b_5}\;
\end{eqnarray}
It is also convenient to define the
ten--form potential $C_{10}$ with the field strength
$R_{11}=dC_{10}- H_3\wedge C_{8}$ which is nonzero due to the
presence of the fermionic directions only,
\begin{eqnarray}
\label{R11=str}
 R_{11} &=&  2i e^{-{\Phi}} E^{\alpha 2} \wedge E^{\beta 1}
 \wedge \bar{\sigma}^{(9)}_{\alpha\beta} -
  e^{-{\Phi}} \left(E^2\wedge \bar{\sigma}^{(10)}\nabla_1\Phi
+ E^1\wedge \bar{\sigma}^{(10)}\nabla_2\Phi\right) = \qquad \nonumber \\
&=&  2i e^{-{\Phi}} E_{\underline{a}}^{\wedge
9}{\sigma}^{{\underline{a}}}_{\alpha\beta} \wedge E^{\alpha 2}
\wedge E^{\beta 1}
 -
  e^{-{\Phi}} E^{\wedge
10} \wedge \left(E^2\nabla_1\Phi + E^1\nabla_2\Phi\right)\; .
\end{eqnarray}

\setcounter{section}{0}
\renewcommand{\thesection}{}
\section{Appendix C. ${SO(1,9)\over  SO(1,7)\times SO(2)}$ moving frame variables (also called Lorentz harmonics)}
\renewcommand{\theequation}{C.\arabic{equation}}
\setcounter{equation}{0}
\renewcommand{\thesection}{C}
\setcounter{subsection}{0}

The moving frame variables (Lorentz harmonics) suitable for the description of the D7--branes
and other 7--branes can be defined as $SO(1,8)\times SO(2)$
covariant blocks of the $SO(1,10)$ valued matrix
\begin{eqnarray}\label{UinSO}
U_{\underline{a}}^{(\underline{b})} := \left({ u_{\underline{a}}
^{\; b} \; , \; {1\over 2} (u_{\underline{a}} ^{\; z} +
\bar{u}_{\underline{a}} ^{\; \bar{z}})\; , \; {1\over 2i}
(u_{\underline{a}} ^{\; z} - \bar{u}_{\underline{a}} ^{\; \bar{z}} )
} \right) \qquad \in \qquad SO(1,9) \; , \qquad \\ \label{uz:=}
u_{\underline{a}} ^{\; z} = {u}_{\underline{a}} ^{\; 8} + i
u_{\underline{a}} ^{\; 9} =(\bar{u}_{\underline{a}} ^{\; \bar{z}})^*
\; , \qquad  \underline{a}, \underline{b}, \underline{c}=0,\, 1\, ,
\ldots \, , \, 9 \;  , \qquad  a,b,c =0,\, 1\, , \ldots \, , \, 7 \;
,
\end{eqnarray}
where  $u_{\underline{a}} ^{\; z} = {u}_{\underline{a}} ^{\; 8} + i
u_{\underline{a}} ^{\; 9} =(\bar{u}_{\underline{a}} ^{\;
\bar{z}})^*$. The condition (\ref{UinSO}) implies
\begin{eqnarray}\label{UTIU=I}
 & U^{\!^T} \eta U=\eta \quad
\Leftrightarrow \quad   \cases{ {u}^{\underline{c} a}
u_{\underline{c}} ^{\; b} =\eta^{ab}\; , \quad {u}^{\underline{a} a}
u_{\underline{a}} ^{\; z} =0= {u}^{\underline{a} a}
\bar{u}_{\underline{a}} ^{\; \bar{z}}\; , \cr  {} \cr \qquad
{u}^{\underline{a} z} u_{\underline{a}} ^{\; z} = 0 =
\bar{u}^{\underline{a} \bar{z}}\bar{u}_{\underline{a}} ^{\;
\bar{z}}\; , \cr  {} \quad \cr
 \qquad {u}^{\underline{a}
z} \bar{u}_{\underline{a}} ^{\; \bar{z}}=-2 \; , }  \qquad
\end{eqnarray}
as well as $U\eta U^{\!^T}=\eta$, which is equivalent to the
following  `unity decomposition'
\begin{eqnarray}\label{UIUT=I}
U\eta U^{\!^T}=\eta \quad \Leftrightarrow \qquad
 & \delta_{\underline{a}}^{\; \underline{b}}= {U}_{\underline{a}}^{(\underline{c})}
U_{(\underline{c})}{}^{\underline{b}} =
 u_{\underline{a}}{}^c {u}_c^{\,\underline{b}}
 - {1\over 2 } {u}_{\underline{a}}^{\; z} \bar{u}^{\underline{b}\bar{z}}\;
  -  {1\over 2 }  \bar{u}_{\underline{a}} ^{\; \bar{z}}\;
  u^{\underline{b}\,  z}  \; .
\end{eqnarray}

The  pseudo-real {\it spinor moving frame} matrix $V_\alpha{}^q$
obeys
\begin{eqnarray}\label{VinSpin}
(V_\alpha{}^q)^*= \gamma^0_{qp} V_\alpha{}^p\; , \qquad
V\sigma^{\underline{a}}V^T= \sigma^{(\underline{b})}
U_{(\underline{b})}{}^{\!\underline{a}}\; , \qquad
V^T\sigma^{(\underline{a})}V= \sigma^{\underline{b}}
U_{\underline{b}}^{(\underline{a})}\; .
\end{eqnarray}
This implies that it is $Spin(1,9)$ group valued and allows to refer on it as on square root of the moving frame variables.

The spinor moving frame matrix $V_\alpha{}^q$  carries one $SO(1,9)$ and one $SO(1,7)$ spinor index
and, hence, can be used as a 'bridge' (in terminology of \cite{GIKOS}) to
convert the $16$ component D=10 Majorana--Weyl spinor (with the
index denoted by a Greek symbol) into a $16$ component pseudo--real
$SO(1,7)$ spinor (the index of which we denote  by $p$ or $q$). For
instance, the pull--backs of the fermionic supervielbein one forms
to the worldvolume, $\hat{E}^{\alpha 1}= d\hat{Z}^{\underline{M}}
(\xi ) E_{\underline{M}}{}^{\alpha 1}(\hat{Z})$, $\hat{E}^{\alpha
2}= d\hat{Z}^{\underline{M}} (\xi ) E_{\underline{M}}{}^{\alpha
2}(\hat{Z})$, which carry the D=10 MW spinorial indices, can be
established to be in one-to-one correspondence with the pseudo--real
one-forms
\begin{eqnarray}\label{EfL1=}
\hat{E}^{q 1}:= \hat{E}^{\alpha 1} V_{\alpha}{}^q=
d\hat{Z}^{\underline{M}} (\xi ) E_{\underline{M}}{}^{\!\!\alpha
1}(\hat{Z}(\xi)) V_{\alpha}{}^q(\xi)\; , \qquad
(\hat{E}^{q 1})^*= \gamma^0_{qp} \hat{E}^{p 1}\; , \qquad \\
\label{EfL2=}
 \hat{E}^{q 2}:= \hat{E}^{\alpha 2} V_{\alpha}{}^q= d\hat{Z}^{\underline{M}}
(\xi ) E_{\underline{M}}{}^{\!\!\alpha 2}(\hat{Z}(\xi))
V_{\alpha}{}^q(\xi)\; , \qquad (\hat{E}^{q 2})^*= \gamma^0_{qp}
\hat{E}^{p 2}\; . \qquad
\end{eqnarray}

Eqs. (\ref{VinSpin}) can be split as
\begin{eqnarray}\label{VgVT=su}
&& V_\alpha \gamma^a V_\beta :=V_\alpha{}^q \gamma^a_{qp}
V_\beta{}^p = \sigma^{\underline{b}}_{\alpha\beta}
u_{\underline{b}}{}^{\!{a}} \; , \qquad \nonumber
\\ &&  V_\alpha{}^q (\delta+ i\gamma^9)_{qp}  V_\beta{}^p=
\sigma^{\underline{b}}_{\alpha\beta} u_{\underline{b}}{}^{\!{z}}\; ,
\qquad V_\alpha{}^q (\delta- i\gamma^9)_{qp}  V_\beta{}^p=
\sigma^{\underline{b}}_{\alpha\beta}
\bar{u}_{\underline{b}}{}^{\!\bar{z}}\; , \qquad
\end{eqnarray}
\begin{eqnarray}
\label{tsu=} && \tilde{\sigma}^{\underline{b}\alpha\beta}
u_{\underline{b}}{}^{\!{a}}= V^\alpha \gamma^a V^\beta :=V_q^\alpha
\gamma^a_{qp} V_p^\beta \; , \qquad \nonumber
\\ && \tilde{\sigma}^{\underline{b}\alpha\beta}  u_{\underline{b}}{}^{\!{z}}= -
V_q^\alpha (\delta- i\gamma^9)_{qp} V_p^\beta \; , \qquad
\tilde{\sigma}^{\underline{b}\alpha\beta}
\bar{u}_{\underline{b}}{}^{\!\bar{z}}= - V_q^\alpha (\delta+
i\gamma^9)_{qp} V_p^\beta \; , \qquad
\end{eqnarray}
\begin{eqnarray}
 \label{VTsV=ug} &&
V_q{\sigma}^{\underline{a}}V_p = \gamma^b_{qp}
u_{b}{}^{\!\underline{a}} - {1\over 2} (\delta+ i\gamma^9)_{qp}
\bar{u}^{\underline{a}\bar{z}} - {1\over 2} (\delta- i\gamma^9)_{qp}
u^{\underline{a} \, {z}}\; .
\end{eqnarray}

To calculate in an easy manner the derivatives of the moving frame variables, one should take into account  that the space tangent to a group manifold
of a Lie group is isomorphic to the Lie algebra of this Lie group.
As far as moving frame variables form the SO(1,9) valued matrix and the spinor moving frame matrix takes its values in  Spin(1,9), doubly covering SO(1,9), this allows us to express the derivatives/variations of both moving frame and spinor moving frame variables (vector
and spinor Lorentz harmonics) in terms of the same Cartan forms,
 \begin{eqnarray}
 \label{dU=UUdU} dU_{\underline{a}}^{(\underline{b})} =
U_{\underline{a}(\underline{c})}
(U^TdU)^{(\underline{c})(\underline{b})}\; , \; \qquad \delta
U_{\underline{a}}^{(\underline{b})} =
U_{\underline{a}(\underline{c})} (U^T\delta
U)^{(\underline{c})(\underline{b})}\; , \; \qquad
\end{eqnarray}\begin{eqnarray}
 \label{dV=VUdU} dV_{\alpha}{}^{q} =
{1\over 4}V_{\alpha}{}^{p} (U^TdU)^{(\underline{c})(\underline{b})}
\sigma_{(\underline{c})(\underline{b})}{}_{pq} \; , \; \qquad \delta
V_{\alpha}{}^{q} = {1\over 4}V_{\alpha}{}^{p} (U^T\delta
U)^{(\underline{c})(\underline{b})}
\sigma_{(\underline{c})(\underline{b})}{}_{pq} \; .  \; \qquad
\end{eqnarray}
When the theory in curved superspace is considered, to keep the
local Lorentz invariance preserved, it is convenient to work with the
covariant generalizations of the Cartan forms, defined with the use
of $SO(1,9)$ covariant derivative $d+w$ instead of the usual
derivative $d$ (see \cite{Dima99,B00}). In our case these
generalized Cartan forms read
 \begin{eqnarray}
 \label{Omz=UDU}
& \Omega^{az}:= u^{\underline{b} a} (du_{\underline{b}}{}^{z} +
w_{\underline{b}}{}^{\underline{c}} u_{\underline{c}}{}^{z}) =:
(udu)^{az} + w^{az} \; , \qquad \\  \label{bOmbz=UDU} &
\bar{\Omega}^{a\bar{z}}:= {u}^{\underline{b} a}
(d\bar{u}_{\underline{b}}{}^{\bar{z}} +
w_{\underline{b}}{}^{\underline{c}}
\bar{u}_{\underline{c}}{}^{\bar{z}})=: (udu)^{a\bar{z}} +
w^{a\bar{z}}  \; , \qquad
\\ \label{A=UDU}
& A := {i\over 2} u^{\underline{b} z}
(d\bar{u}_{\underline{b}}{}^{\bar{z}} +
w_{\underline{b}}{}^{\underline{c}}
\bar{u}_{\underline{c}}{}^{\bar{z}}) =: {i\over 2}
((ud{u})^{z\bar{z}} + w^{z\bar{z}})\; , \quad \\  \label{om=UDU}  &
\omega^{ab}:= u^{\underline{c} a} (du_{\underline{c}}{}^{b} +
w_{\underline{c}}{}^{\underline{c}^\prime}
u_{\underline{c}^\prime}{}^{b})=: (udu)^{ab}+ w^{ab}\; .  \qquad
\end{eqnarray}
These definitions can be collected in the expression for the
$SO(1,9)\otimes SO(1,7)\otimes SO(2)$ covariant derivatives  of the
moving frame vectors
\begin{eqnarray}
 \label{Duz:==}
&&  Du_{\underline{b}}{}^{z}:= du_{\underline{b}}{}^{z} - iA
u_{\underline{b}}{}^{z} + w_{\underline{b}}{}^{\underline{c}}
u_{\underline{c}}{}^{z} = {u}_{\underline{b} a} \Omega^{az}\; ,
\qquad \\  \label{Dubz:==}
 && D\bar{u}_{\underline{b}}{}^{\bar{z}}
 :=
d\bar{u}_{\underline{b}}{}^{\bar{z}} + i A
\bar{u}_{\underline{b}}{}^{\bar{z}} +
w_{\underline{b}}{}^{\underline{c}}
\bar{u}_{\underline{c}}{}^{\bar{z}}= {u}_{\underline{b} a}
\bar{\Omega}^{a\bar{z}} \; , \qquad
\\ \label{Dua:==}
&& Du_{\underline{b}}{}^{a}:= du_{\underline{b}}{}^{a} -
u_{\underline{b}}{}^{b}\omega_b{}^c +
w_{\underline{b}}{}^{\underline{c}} u_{\underline{c}}{}^{z} =
{1\over 2} \bar{u}_{\underline{b}}{}^{\bar{z}} \Omega^{az} +
{1\over 2}  u_{\underline{b}}{}^{z} \bar{\Omega}^{a\bar{z}}  \; .
\qquad
\end{eqnarray}
The covariant version of (\ref{dV=VUdU}) can be, in its turn,
written as an expression for the $SO(1,9)\otimes SO(1,7)\otimes
SO(2)$ covariant derivatives of the spinor moving frame variables,
{\it i.e} of the spinor moving frame matrix,
\begin{eqnarray}
 \label{DV:==}
 DV_{\alpha}{}^{q} &:=& dV_{\alpha}{}^{q}  + {1\over 4} w^{\underline{a}\underline{b}}
\sigma_{\underline{a}\underline{b}}{}_{\alpha}{}^{\beta}
V_{\beta}{}^{q} - {1\over 4} \omega^{ab}
V_{\alpha}{}^{p}\gamma_{ab}{}_{pq} - {1\over 2} A
V_{\alpha}{}^{p}(\gamma_9)_{pq} = \qquad \nonumber \\ &=& {1\over 4}
\; \Omega^{az}\; V_{\alpha}{}^{p}(\gamma_a(\delta + i\gamma_9))_{pq}
+ {1\over 4} \; \bar{\Omega}^{a\bar{z}} \;
V_{\alpha}{}^{p}(\gamma_a(\delta - i\gamma_9))_{pq}  \; .  \qquad
\end{eqnarray}

Some other relations useful for our study are
\begin{eqnarray}
\label{V-1tsV-1=ug} & V^q\tilde{\sigma}_{\underline{a}}V^p = u^{\;
b}_{\underline{a}} \gamma_{b\, qp}  + {1\over 2} (\delta +
i\gamma^9)_{qp} u_{\underline{a}}{}^{\!{z}}+ {1\over 2} (\delta-
i\gamma^9)_{qp} \bar{u}_{\underline{a}}{}^{\!\bar{z}}\; .
\end{eqnarray}
({\it cf.} Eqs. (\ref{VTsV=ug}), (\ref{d8Iden})),
\begin{eqnarray} & V_q{\sigma}^{\underline{a}\underline{b}}V^p =
u_a^{\underline{a}}u_b^{\underline{b}}\gamma^{ab}_{qp} +
u_a^{[\underline{a}}u^{\underline{b}]z} (\gamma^a(\delta -
i\gamma^9))_{qp}+
u_a^{[\underline{a}}\bar{u}{}^{\underline{b}]\bar{z}}
(\gamma^a(\delta - i\gamma^9))_{qp} - i
{u}^{z[\underline{a}}\bar{u}{}^{\underline{b}]\bar{z}}
i\gamma^9_{qp}\; , \quad \nonumber \\
\label{VsVVstV=} & V_p{\sigma}^{\underline{a}}V_{p^\prime}\;
V^q\tilde{\sigma}_{\underline{a}}V^{q^\prime} =
\gamma^b_{pp^\prime}\, \gamma_{b\; qq^\prime} + {1\over 2} (\delta+
i\gamma^9)_{pp^\prime} (\delta+ i\gamma^9)_{qq^\prime} + {1\over 2}
(\delta- i\gamma^9)_{pp^\prime} (\delta - i\gamma^9)_{qq^\prime} \;
. \quad
\end{eqnarray}
Notice the difference of this latter relation with
\begin{eqnarray}
\label{VsVVsV=} & V_p{\sigma}^{\underline{a}}V_{p^\prime}\; V_{q}
{\sigma}_{\underline{a}}V_{q^\prime} = \gamma^b_{pp^\prime}\,
\gamma_{b\; qq^\prime} - {1\over 2} (\delta+ i\gamma^9)_{pp^\prime}
(\delta- i\gamma^9)_{qq^\prime} - {1\over 2} (\delta-
i\gamma^9)_{pp^\prime} (\delta + i\gamma^9)_{qq^\prime} \; .
\end{eqnarray}
This relation can be used to find that the famous $D=10$ Fierz
identity (\ref{10Did=}),
$\sigma_{\underline{a}\,(\alpha\beta}\sigma_{\gamma)
\delta}^{\underline{a}}=0$, is represented by Eq. (\ref{d8Iden}),
\begin{eqnarray}
\label{theFI=} & \gamma_{b(pp^\prime}\, \gamma^b_{q)q^\prime} =
{1\over 2} (\delta+ i\gamma^9)_{(pp^\prime} (\delta-
i\gamma^9)_{q)q^\prime} + {1\over 2} (\delta-
i\gamma^9)_{(pp^\prime} (\delta + i\gamma^9)_{q)q^\prime} \; .
\end{eqnarray}

To work with fermionic torsion (see Eq. (\ref{Deq=}) below) one uses
the spin-tensor
\begin{eqnarray}
\label{fppqq:=}
 && \mathbf{f}_{pp^\prime}{}^{qq^\prime}  :=  \delta_{(p}{}^{q}\delta_{p^\prime )}{}^{q^\prime} - {1\over
 2} V_p{\sigma}^{\underline{a}}V_{p^\prime}\;
V^q\tilde{\sigma}_{\underline{a}}V^{q^\prime} = \qquad \nonumber \\
 && \quad = \delta_{(p}{}^{q}\delta_{p^\prime )}{}^{q^\prime}
 - {1\over 2} \gamma^b_{pp^\prime}\gamma_b^{qq^\prime} -
  {(\delta + i\gamma^9)_{pp^\prime} \over 2}
  {(\delta + i\gamma^9)_{qq^\prime} \over 2}-
  {(\delta - i\gamma^9)_{pp^\prime}\over 2}
  {(\delta - i\gamma^9)_{qq^\prime}\over 2}\; . \qquad
\end{eqnarray}
One can easily check that its trace in 'lower' indices, $\mathbf{f}_{pp}{}^{qq^\prime}:= \delta_{pp^\prime}\mathbf{f}_{pp^\prime}{}^{qq^\prime}$ is proportional to unity matrix $\delta^{qq^\prime}$,
\begin{eqnarray}
\label{trfqq:=}
&& \mathbf{f}_{pp}{}^{qq^\prime}  = - 7 \delta^{qq^\prime} \; , \qquad
\end{eqnarray}
and also that, as a consequence of Eqs. (\ref{Tal1=str}),
(\ref{Tal2=str}) or (\ref{Tfff=str}) above,
\begin{eqnarray}
\label{VTV=}
  V_p{}^\alpha V_{p^\prime}{}^\beta T_{\alpha 1\, \beta 1}{}^{\gamma 1}
 V_{\gamma}{}^{q} = - 2 \mathbf{f}_{pp^\prime}{}^{qq^\prime} V_{q^\prime}{}^\delta D_{\delta 1}
 e^{-\Phi} \; , \qquad \nonumber \\ V_p{}^\alpha V_{p^\prime}{}^\beta T_{\alpha 2\, \beta 2}{}^{\gamma 2}
 V_{\gamma}{}^{q} = - 2 \mathbf{f}_{pp^\prime}{}^{qq^\prime} V_{q^\prime}{}^\delta D_{\delta 2}
 e^{-\Phi} \; . \qquad
\end{eqnarray}

Important properties of the
$\mathbf{f}_{pp^\prime}{}^{qq^\prime}$ spin-tensor are
\begin{eqnarray}
\label{f(1+ig)3s=}
 (\delta + i\gamma^9)_{(q_1}{}^{q^\prime}(\delta + i\gamma^9)_{q_2}{}^{p} (\delta + i\gamma^9)_{q_3)}{}^{p^\prime}
 \mathbf{f}_{pp^\prime}{}^{qq^\prime} =0 \; , \qquad \nonumber \\
(\delta - i\gamma^9)_{(q_1}{}^{q^\prime}(\delta -
i\gamma^9)_{q_2}{}^{p} (\delta - i\gamma^9)_{q_3)}{}^{p^\prime}
 \mathbf{f}_{pp^\prime}{}^{qq^\prime} =0 \; . \qquad
\end{eqnarray}

\bigskip
\setcounter{section}{0}
\renewcommand{\thesection}{}
\section{Appendix D.  Induced worldvolume superspace geometry for 7--branes}
\renewcommand{\thesection}{D}
\renewcommand{\theequation}{D.\arabic{equation}}
\setcounter{equation}{0} \setcounter{subsection}{0}

The induced worldvolume geometry of the 7--brane worldvolume
superspace embedded into the D=10 type IIB superspace in such a way
that the supermebedding equation (\ref{SembEq}) and the conventional
constraints  hold (all these can be collected in Eqs.
(\ref{Eua=eau}) and  (\ref{Du=})), is characterized by the bosonic
torsion
\begin{eqnarray}
\label{Dea=} De^a= -i e^q\wedge e^p (\gamma^a + h\gamma^ah^T)_{pq} +
2i e^b\wedge e^q (h\gamma^a\chi_b)_q + i e^c\wedge e^b
\chi_b\gamma^a\chi_c \;
\end{eqnarray}
and fermionic torsion two--form,
\begin{eqnarray}
\label{Deq=}  &De^q &=  {1\over 4} e^p \wedge \Omega^{bz}
(\gamma_b(\delta + i\gamma^9))_{pq} + {1\over 4} e^p \wedge
\bar{\Omega}^{b\bar{z}} (\gamma_b(\delta - i\gamma^9))_{pq} - \qquad
\nonumber \\ && - e^p\wedge e^{p^\prime}
\mathbf{f}_{pp^\prime}{}^{qq^\prime}
V_{q^\prime}{}^\alpha D_{\alpha 1}e^{-\Phi} + \qquad \nonumber \\
&&+ e^a\wedge e^p u_a{}^{\underline{b}} (V_p{}^\alpha T_{\alpha 1 \,
\underline{b}}{}^{\beta 1} V_\beta{}^q + h_p{}^{p^\prime }
V_{p^\prime}{}^\alpha T_{\alpha 2 \, \underline{b}}{}^{\beta 1}
V_\beta{}^q ) +{1\over 4} e^{b^\prime}\wedge e^{a^\prime}
u_{a^\prime }{}^{\underline{a}}u_{b^\prime }{}^{\underline{b}}
\hat{T}_{\underline{a}\underline{b}}{}^{\alpha 1} V_\alpha{}^q = \qquad \nonumber \\
&=&  - {i\over 2} e^p\wedge e^{p^\prime}\left((h(\delta +
i\gamma^9)\chi_b)_{(p} \; (\gamma^b(\delta + i\gamma^9))_{p^\prime
)q} + c.c. - 2i \mathbf{f}_{pp^\prime}{}^{qq^\prime}
V_{q^\prime}{}^\alpha D_{\alpha 1}e^{-\Phi}
  \right) - \qquad \nonumber \\
&&  -{1\over 4}  e^a\wedge  e^p \;  \left(K_{ab}{}^z- i
\chi_a(\delta +
i\gamma^9)\chi_b \right) (\gamma^b(\delta + i\gamma^9))_{pq} - \qquad \nonumber \\
&&  -{1\over 4} e^a\wedge  e^p \;  \left(\bar{K}_{ab}{}^{\bar{z}}- i
\chi_a(\delta - i\gamma^9)\chi_b
\right) (\gamma^b(\delta - i\gamma^9))_{pq}   + \qquad \nonumber \\
&& +   e^a\wedge  e^p \;  u_{a}{}^{\underline{b}} \left(
V_p{}^\alpha T_{\alpha 1 \, \underline{b}}{}^{\beta 1} V_\beta{}^q +
h_p{}^{p^\prime } V_{p^\prime}{}^\alpha T_{\alpha 2 \,
\underline{b}}{}^{\beta 1} V_\beta{}^q \right)  + {1\over 4}
e^{b^\prime}\wedge e^{a^\prime} u_{a^\prime
}{}^{\underline{a}}u_{b^\prime }{}^{\underline{b}}
\hat{T}_{\underline{a}\underline{b}}{}^{\alpha 1} V_\alpha{}^q \; ,
\end{eqnarray}
where $ \mathbf{f}_{pp^\prime}{}^{qq^\prime}$ is defined in Eq.
(\ref{fppqq:=}).

In particular, the dimension $1/2$ fermionic torsion, according to
Eq. (\ref{Deq=}) reads
\begin{eqnarray}
\label{Tfff(w-sh)=}
 T_{pp^\prime}{}^q= - i \left((h(\delta +
i\gamma^9)\chi_b)_{(p} \; (\gamma^b(\delta + i\gamma^9))_{p^\prime
)q} + (h(\delta - i\gamma^9)\chi_b)_{(p} \; (\gamma^b(\delta -
i\gamma^9))_{p^\prime )q} - \right. \; \nonumber \\ \left.  - 2i
\mathbf{f}_{pp^\prime}{}^{qq^\prime} V_{q^\prime}{}^\alpha D_{\alpha
1}e^{-\Phi}
  \right)\; , \qquad
\end{eqnarray}
%where $ \mathbf{f}_{pp^\prime}{}^{qq^\prime}$ is defined in Eq. (\ref{fppqq:=}).
 and the dimension $3/2$ fermionic torsion
spin-tensor is
\begin{eqnarray}
\label{Tbff(w-sh)=}
 T_{ap}{}^q& =& {1\over 4} \;  \left(K_{ab}{}^z- i \chi_a(\delta +
i\gamma^9)\chi_b \right) (\gamma^b(\delta + i\gamma^9))_{pq} + \qquad \nonumber \\
&&  + {1\over 4}  \; \left(\bar{K}_{ab}{}^{\bar{z}}- i \chi_a(\delta
- i\gamma^9)\chi_b
\right) (\gamma^b(\delta - i\gamma^9))_{pq}   -  \qquad \nonumber \\
&& \quad -  u_{a }{}^{\underline{b}} \left( V_p{}^\alpha T_{\alpha 1
\, \underline{b}}{}^{\beta 1} V_\beta{}^q + h_p{}^{p^\prime }
V_{p^\prime}{}^\alpha T_{\alpha 2 \, \underline{b}}{}^{\beta 1}
V_\beta{}^q \right)\; . \qquad
\end{eqnarray}
Notice that
\begin{eqnarray}
\label{Tbffga=Kab}
 && T_{ap}{}^q\gamma_{b}^{pq} =4 \left(K_{ab}{}^z + \bar{K}_{ab}{}^{\bar{z}} - 2i \chi_a\chi_b \right)
-  u_{a }{}^{\underline{c}} \left(  (V\gamma_{b}V)_\beta{}^\alpha
 \hat{T}_{\alpha 1 \, \underline{c}}{}^{\beta 1} +
(V\gamma_{b}hV)_\beta{}^\alpha \hat{T}_{\alpha 2 \,
\underline{c}}{}^{\beta 1} \right)\; ,  \qquad \nonumber \\
&& T_{ap}{}^q(i\gamma_{b}\gamma^{(9)})^{pq} = -4 \left(K_{ab}{}^z -
\bar{K}_{ab}{}^{\bar{z}} +2 \chi_a\gamma^{(9)}\chi_b \right) - i
u_{a }{}^{\underline{c}} \left(
(V\gamma_{b}\gamma^{(9)}V)_\beta{}^\alpha
 \hat{T}_{\alpha 1 \, \underline{c}}{}^{\beta 1} + \right.  \qquad \nonumber \\ && \left. \hspace{5cm} +
(V\gamma_{b}\gamma^{(9)}hV)_\beta{}^\alpha \hat{T}_{\alpha 2 \,
\underline{c}}{}^{\beta 1} \right)\; ,
\end{eqnarray}
which implies that the D$7$--brane scalar field equation can be
formulated as expressions for the chiral gamma-traces,
$T_{ap}{}^q\gamma^{a}(\delta \pm i\gamma^9)^{pq}$,  of the dimension
1 torsion superfield $T_{ap}{}^q$ (see Sec. 3.3.2 for a discussion).

Using (\ref{f(1+ig)3s=}) and the identities $(\delta + i\gamma^9)
\gamma_b(\delta + i\gamma^9)=0=(\delta - i\gamma^9)
\gamma_b(\delta - i\gamma^9)$ one finds
\begin{eqnarray} \label{T(1+ig)3s=}
 (\delta + i\gamma^9)_{(q_1}{}^{q}(\delta + i\gamma^9)_{q_2}{}^{p} (\delta + i\gamma^9)_{q_3)}{}^{p^\prime}
 T_{pp^\prime}{}^{q} =0 \; , \qquad \nonumber \\
 (\delta - i\gamma^9)_{(q_1}{}^{q}(\delta - i\gamma^9)_{q_2}{}^{p} (\delta - i\gamma^9)_{q_3)}{}^{p^\prime}
 T_{pp^\prime}{}^{q} =0\; . \qquad
\end{eqnarray}
%(\ref{T(1+ig)3s=})
Similarly
\begin{eqnarray} \label{Tfbf(1+ig)2=}
 (\delta + i\gamma^9)_{q_1}{}^{q}(\delta + i\gamma^9)_{q_2}{}^{p}
 T_{pb}{}^{q} =0 \; , \qquad \nonumber \\
 (\delta - i\gamma^9)_{q_1}{}^{q}(\delta - i\gamma^9)_{q_2}{}^{p}
 T_{pb}{}^{q} =0 \; . \qquad
\end{eqnarray}
Furthermore,
\begin{eqnarray} \label{TfffPY+cc=}
&& {1\over 2} T_{(q_1q_2}{}^p ((\delta + i\gamma^9)_{q_3)p}\Upsilon
+ c.c.) = i\gamma^{b}_{(q_1q_2} (h(\delta +
i\gamma^9)\chi_b)_{q_3)}\Upsilon + c.c. + \nonumber \qquad \\ &&
\qquad + {1\over 2} \gamma^{b}_{(q_1q_2} ((\delta +
i\gamma^9)\gamma_bV\widehat{D_1e^{-\Phi}})_{q_3)}\Upsilon - {1\over
2} (\delta + i\gamma^9)_{(q_1q_2} ((\delta -
i\gamma^9)V\widehat{D_1e^{-\Phi}})_{q_3)}\Upsilon + c.c. \; . \qquad
\end{eqnarray}

Eqs. (\ref{Dea=}), (\ref{Deq=}) follow from the conventional
constraints (\ref{Eua=ea}), (\ref{Eq1=eq}). The fermionic field
$\chi$, appearing in the decomposition of the pull--back of the
second fermionic supervielbein, Eq. (\ref{Eq2=}), is related with
the spinorial derivative of the $SO(1,7)$ spin-tensor $h_p{}^q$
(\ref{h8=}), appearing in the same Eq. (\ref{Eq2=}), by
\begin{eqnarray}
\label{Dphpq=} D_{(p}h_{p^\prime )}{}^{q} &= & - i (\gamma^a +
h\gamma^a h^T)_{pp^\prime} \chi_a{}^q + {i\over 2}
(h(\delta+i\gamma^9)\chi_a)_{(p}\,
(h\gamma^a(\delta+i\gamma^9))_{p^\prime )q}  + \qquad \nonumber
\\ & + &{i\over 2} (h(\delta-i\gamma^9)\chi_a)_{(p}\,
(h\gamma^a(\delta-i\gamma^9))_{p^\prime )q} + \qquad \nonumber
\\ & + & 2 \mathbf{f}_{pp^\prime}{}^{qq^\prime} (hV)_{q^\prime}{}^\alpha
D_{\alpha 1}e^{-\Phi} - 2 (h\otimes h \cdot
\mathbf{f})_{pp^\prime}{}^{qq^\prime} V_{q^\prime}{}^\alpha
D_{\alpha 2}e^{-\Phi}\; . \qquad
\end{eqnarray}
Eq. (\ref{Dphpq=}) can be obtained from the $\propto e^p \wedge
e^{p^\prime }$ component of the integrability condition $D(E^{\alpha
2} V_\alpha{}^q - e^ph_p{}^q -e^a\chi_a{}^q)=0$ of the conventional
fermionic superembedding condition (\ref{Eq2=eh+}). A simple, but important consequence of Eq.
(\ref{Dphpq=}) is given by Eq. (\ref{trDphpq=}), 
\begin{eqnarray}
\label{trDphpq=A} D_{p}h_{p}{}^{q} &= & -14\left( (hV)_{q}{}^\alpha
D_{\alpha 1}e^{-\Phi} + V_{q^\prime}{}^\alpha
D_{\alpha 2}e^{-\Phi}\right) \; . \qquad
\end{eqnarray}
To derive it Eqs. (\ref{trfqq:=}) and (\ref{hhT=-1})  should be taken into account.

\bigskip
\setcounter{section}{0}
\renewcommand{\thesection}{}
\section{Appendix E.  Derivation of and complete form of some equations}
\renewcommand{\thesection}{E}
\renewcommand{\theequation}{E.\arabic{equation}}
\setcounter{equation}{0} \setcounter{subsection}{0}

\subsection*{E1.  Derivation of Eqs. (\ref{DY=gW+}) and
(\ref{cUbp=0})}

Here we present some detail on solving Eq. (\ref{QdG(3/2)=0}). It is
convenient to begin by contracting the indices $q_1q_2q_3$ with
different  sets of three projectors $(\delta \pm i\gamma^9)$. In
particular, using the identity $(\delta + i\gamma^9)\gamma_b(\delta
+ i\gamma^9)=0= (\delta - i\gamma^9)\gamma_b(\delta - i\gamma^9)$
and Eqs. (\ref{T(1+ig)3s=}), one finds that the contraction with
three copies of the same projector produces very simple equations
\begin{eqnarray}
\label{(1+g)3DY=0E}
 (\delta + i\gamma^9)_{(q_1q_2}  \, ((\delta +
i\gamma^9)D \Upsilon)_{q_3)} =0 \; , \qquad   (\delta -
i\gamma^9)_{(q_1q_2} \, ((\delta - i\gamma^9)D
\bar{\Upsilon})_{q_3)} =0 \;
 \qquad
\end{eqnarray}
which  imply
\begin{eqnarray}
\label{(1+i9)DY=0E} (\delta + i\gamma^9)_{qp}D_p \Upsilon = 0\; , \qquad (\delta -
i\gamma^9)_{qp} D_p \bar{\Upsilon} = 0\; . \qquad
 \end{eqnarray}
These equations are solved by
  \begin{eqnarray}
\label{DY=nl} D_q \Upsilon = - 2i (\delta - i\gamma^9)_{qp}\tilde{{\cal W}}{}^p\; ,
\qquad D_q \bar{\Upsilon} = - 2i (\delta + i\gamma^9)_{qp}\tilde{{\cal W}}{}^p\;
\end{eqnarray}
with some fermionic field $\tilde{{\cal W}}{}^p$ whose relation the ${{\cal W}}{}^p$
superfields of Eqs. (\ref{dA-B-C=}) is to be determined.

Now, multiplying (\ref{QdG(3/2)=0}) by two (but not three as above) $(\delta +i\gamma^9)$ projectors  and using Eqs. (\ref{theFI=}), (\ref{Tfff(w-sh)=}) and
$(\eta-F)^{-1}=\eta +
F(\eta-F)^{-1}$,   after some algebra one arrives at
\begin{eqnarray}\label{QdG(3/2)=02}
 & (\delta + i\gamma^9)_{q_1q_2}  (\delta -
i\gamma^9)_{q_3p}\left(\tilde{{\cal W}}{}^p-{\cal W}^p+ \Upsilon \Lambda_{1 p}\right) = (\gamma^b (\delta + i\gamma^9))_{q_3(q_1}  (\delta +
i\gamma^9)_{q_2)p} \; {\cal U}^p_b \; , \qquad
\end{eqnarray} where
\begin{eqnarray} \label{cUbp:=}
& {\cal U}^p_b:= {\cal U}^p_b ({\cal W}, F_{cd},\Upsilon)\;= (F(\eta
-F)^{-1})_b{}^{c} (\gamma_c{\cal W})_{p}
 - 2 \Upsilon \Lambda_{1 p} - \bar{\Upsilon} (h\chi_b)_p\,
\;  \qquad
\end{eqnarray}
and $\Lambda^1_{q}$ is defined in  (\ref{L1=def}).  Notice that  ${\cal U}^p_b:= {\cal U}^p_b ({\cal W}, F_{cd},\Upsilon)$ in Eq.
(\ref{cUbp:=}) and, hence, the {\it r.h.s.} of Eq. (\ref{QdG(3/2)=02}) does not depend
on $\tilde{{\cal W}}^p$.

Let us discuss the decomposition of Eq. (\ref{QdG(3/2)=02}) on the
irreducible $SO(1,7)$ representations with respect to the
symmetrized pair of indices $(q_1q_2)$. The list of symmetric
$16\times 16$ matrices is given in Eq. (\ref{SYMbasis}) of Appendix
A.  Only one of the  irreducible parts,  $\propto (\delta + i
\gamma^9)_{q_1q_2} $ (which we denote by ${\mathbf 1}$),  contains
$\tilde{{\cal W}}{}^p$ and can be used to determine its form, which
we are going to discuss below. Other parts do not contain that and,
hence, can (and really do) put restrictions on ${\cal W}$, $F_{ab}$
and $\Upsilon$.

Indeed, although one can see that the  $\bar{\mathbf 1}$, ${\mathbf 8}$ and
${\mathbf{56}}$ irreducible parts of Eq. (\ref{QdG(3/2)=02})  are satisfied identically
\footnote{The proof is  basically reduced to the observations that $\gamma^a(\delta + i
\gamma^9)=(\delta - i \gamma^9)\gamma^a$, $(\gamma^{abc}\gamma^9)(\delta + i
\gamma^9)=(\delta - i \gamma^9)(\gamma^{abc}\gamma^9)$ and $(\delta\pm i
\gamma^9)(\delta\mp i \gamma^9)=0$.}, the ${\mathbf{70}}$ irreducible part implies
\begin{eqnarray}\label{QdG(3/2)=70}
\gamma^b\gamma^{a_1a_2a_3a_4} (\delta + i\gamma^9){\cal U}_b =0 \; . \qquad
\end{eqnarray}
Eq. (\ref{QdG(3/2)=70}) has only trivial solutions,
\begin{eqnarray}\label{cUbp=00} (\delta + i\gamma^9){\cal U}_b =0\; .
\end{eqnarray}
Taking into account the explicit form of ${\cal U}_b$, Eq. (\ref{cUbp:=}), one sees that
(\ref{cUbp=00}) coincides with  Eq. (\ref{cUbp=0}).

Now, Eq. (\ref{QdG(3/2)=02}) with ${\cal U}_b=0$  gives $\tilde{\cal W}{}^p={\cal W}{}^p  -\Upsilon \Lambda^1_{p}$,
so that Eq.  (\ref{DY=gW+}) is valid,
\begin{eqnarray}
\label{DY=gW+E} & D_q \Upsilon = - 2i (\delta -
i\gamma^9)_{qp}\left({\cal W}{}^p  -\Upsilon \Lambda^1_{p}
\right)\; . \qquad
\end{eqnarray}

\subsection*{E2.  More complete form of Eqs. (\ref{DcW=ag+}) and (\ref{1ta=1FF})}

A more complete form of Eq. (\ref{DcW=ag+}) is given by
\begin{eqnarray} \label{DcW=ag+1E} & D_p{\cal W}^{q^\prime}=
ia_{ab}\gamma^{ab}_{pq^\prime} +
i\tilde{a}_{ab}(\gamma^{ab}\gamma^9)_{pq^\prime} - \nonumber \\ {}&
- {1\over 2} (\gamma_{a}(\delta - i\gamma^9))_{pq^\prime}(\eta -
F)^{-1\; ab}
 D_b{\Upsilon} -   {1\over 2} (\gamma_{a}(\delta + i\gamma^9))_{pq^\prime}(\eta - F)^{-1\; ab}
 D_b\bar{\Upsilon}  + \nonumber \\ {}& + {i\over 16}
(\gamma_{a}(\delta - i\gamma^9))_{pq^\prime} \; ({\cal W}\gamma^b
\gamma^a(\delta - i \gamma^9)h\chi_b)+ {i\over 16}
(\gamma_{a}(\delta + i\gamma^9))_{pq^\prime} \; ({\cal W}\gamma^b
\gamma^a(\delta + i \gamma^9)h\chi_b) + \nonumber \\  {}&  + {i\over
96} (\gamma_{abc}(\delta - i\gamma^9))_{pq^\prime} \; ({\cal
W}\gamma^{d} \gamma^{abc}(\delta - i \gamma^9)h\chi_b)+ {i\over 96}
(\gamma_{abc}(\delta + i\gamma^9))_{pq^\prime} \; ({\cal
W}\gamma^{d} \gamma^{abc}(\delta + i \gamma^9)h\chi_b) + \nonumber \\
{}& + {\cal O}({\Lambda}_1)+ {\cal O}( D_p {\Lambda}_1 )\; . \qquad
\end{eqnarray}
The terms denoted by ${\cal O}({\Lambda}_1\, , D_p {\Lambda}_1 )$
contain contributions of the fermionic flux (pull--back of the
background fermionic superfield) and of the second derivative of the
dilaton superfield, which is expressed through bosonic fluxes. These
terms vanish in the case of flat tangent type IIB superspace.

A more complete form of Eq. (\ref{1ta=1FF}) reads
\begin{eqnarray}\label{1ta=1FF+OY}\nonumber
  \eta_{b[c_1} \tilde{a}_{c_2c_3]} &=&  - {\tilde{q}^\prime \over
\sqrt{|\eta+F|}}(\eta - F)_{b[c_1} F_{c_2c_3]}  - {1\over 2 \cdot 4!}
({\cal W}\gamma_a\gamma_{c_1c_2c_3}\gamma^9h\gamma^a\chi_b) - \nonumber \\ && - {1\over 4\cdot 4!}
(\gamma_{c_1c_2c_3}\gamma^9)_{pq} T_{b\, p}{}^{q^\prime}    \left(
(\delta + i\gamma^9)_{q^\prime q}\Upsilon - (\delta -
i\gamma^9)_{q^\prime q}\bar{\Upsilon} \right) + \qquad \nonumber \\
 &&+ {i\over 8 \cdot 4!}
(\gamma_{c_1c_2c_3}\gamma^9)_{pq} T_{ pq}{}^{q^\prime} (\gamma_b
{\cal W})_{q^\prime} + {\cal O} (\Lambda_1) +  {\cal O}
(D_p\Lambda_1) \; . \qquad  \nonumber \\ {}
\end{eqnarray}
Using the  explicit expressions for $T_{ pq}{}^{q^\prime}$ and
$T_{b\, p}{}^{q^\prime}$ in Eq. (\ref{Tfff(w-sh)=}) and
(\ref{Tbff(w-sh)=}) one can check that the third  and the fourth
terms in the {\it r.h.s} of this equation are equal to zero  in the
case of vanishing background fluxes, in particular, in flat target
superspace. }

\end{document}